\documentclass{kapproc} 
\usepackage{epsfig}
\def\gtrsim{\mathrel{\hbox{\rlap{\hbox{\lower4pt\hbox{$\sim$}}}\hbox{$>$}}}}
\def\lesssim{\mathrel{\hbox{\rlap{\hbox{\lower4pt\hbox{$\sim$}}}\hbox{$<$}}}}
\newcommand{\ga}{\gtrsim}
\newcommand{\la}{\lesssim}

\begin{document}

\articletitle[X-Ray and Radio Observations of Bright GeV Sources]
{X-Ray and Radio Observations of Bright GeV Sources}

\author{Mallory S. E. Roberts}
\affil{Dept of Physics, McGill University\\
3600 University St., Montr\'eal, Qu\'ebec H3A 2T8, Canada}
\email{roberts@physics.mcgill.ca}

\author{Roger W. Romani}
\affil{Dept. of Physics, Stanford University\\
Stanford, Calif. 94305, U.S.A.}

\author{Nobuyuki Kawai}
\affil{Cosmic Radiation Laboratory, The Institute of Physical and 
Chemical Research (RIKEN)\\
2-1 Hirosawa, Wako, Saitama 351-0198, Japan}

\author{Bryan M. Gaensler}
\affil{Center for Space Research, Massachusetts Institute of Technology\\
70 Vassar Street, Cambridge, Mass. 02139, U.S.A.}

\author{Simon Johnston}
\affil{Research Center for Theoretical Astrophysics\\
University of Sydney, NSW 2006, Australia}



\begin{abstract}
We present X-ray and radio studies of sources which are bright
above 1 GeV ($F_{> \rm 1GeV} \ge 4\times 10^{-8} {\rm ph}\, {\rm cm}^{-2}\,
{\rm s}^{-1}$). Only 11 out of $\sim 30$ of these $\gamma-$ray sources
have been identified with lower energy counterparts:  5 blazars and 
6 pulsars. 
Three of these pulsars are surrounded by radio pulsar wind nebulae (PWN), 
two of which are also seen as bright, extended X-ray synchrotron nebulae.  
The ASCA X-ray telescope has observed 28 of the bright GeV sources, revealing 
an excess of
$F_{2-10\rm keV} > 10^{-12} {\rm ergs}\, {\rm cm}^{-2}\ {\rm s}^{-1}$ sources
within the {\it EGRET} error contours of the unidentified sources. 
Although several supernova remnants are positionally coincident with
these sources, we find no X-ray evidence of high energy particle 
production in SNR shell shocks consistent with the GeV positions. 
We also present initial
results from follow on radio imaging studies of several fields containing
unidentified sources.
We have discovered new X-ray/radio nebulae
in three of these fields
which are strong candidates for PWN. These sources, along with a
similar nebula in CTA 1 and the PWN around PSR B1853+01 in W44, are all
positionally coincident with variable {\it EGRET} sources. This suggests
a class of variable $\gamma-$ray sources associated with synchrotron emitting 
regions powered by the winds of young pulsars.

\end{abstract}

\section{Introduction}
\begin{figure}[ht]
\centerline{\epsfig{file=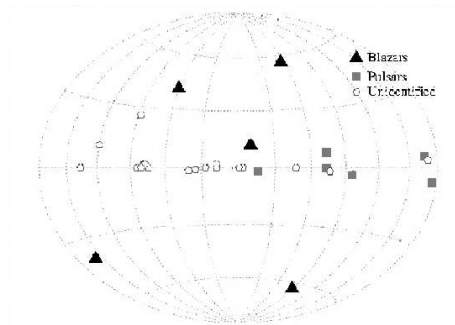, width=0.5\hsize}}
\caption{\protect\inx{The } Galactic 
distribution of bright GeV sources on the sky
(\cite{lm97}).}
\label{gal_dist}
\end{figure}

Sources that are bright above 1 GeV are an important subset of the sources 
detected by {\it EGRET}. Lamb and Macomb (1997) first noted that the $\sim 30$
sources 
with GeV fluxes $F_{>\rm 1GeV} \ge 4\times 10^{-8} {\rm ph}\, {\rm cm}^{-2}\,{\rm s}^{-1}$
are tightly clustered along the Galactic plane (Figure~\ref{gal_dist}), 
making up the majority
of low-latitude {\it EGRET} sources. 6 of the $\sim8$ 
identified $\gamma-$ray
pulsars are in this set but only a small fraction (5/$\sim 80$) 
of the known blazars are. The unidentified bright GeV sources
are therefore good candidates for further pulsar
identifications. Virtually none of them are likely to be background blazars
seen through the Galaxy. Of special importance to our understanding of
pulsar emission mechanisms is the number of these sources which are
radio-quiet pulsars. Outer-gap models (\cite{r96}) predict the majority of
these sources should be pulsars whose $\gamma-$ray emission is beamed towards 
us, but whose radio emission is not. Polar-cap models (\cite{h00}) 
predict relatively few such pulsars, making other source classes necessary. 
One such proposed class is supernova remnant (SNR) shocks, 
generally assumed to produce the majority of observed cosmic rays 
(eg. \cite{bv00}). 
Models of SNR predict hard, generically flat spectra in the GeV range 
(\cite{b99}). 
Other proposed source classes, such as colliding winds in massive binary 
systems (\cite{eu93}) and isolated accreting black holes (\cite{p99}),
have more tentative predictions of their $\gamma-$ray luminosity at GeV energies.

There are practical observational reasons to focus on the bright GeV sources
among the Galactic source population of $\gamma-$ray sources. 
In general, the point spread function (PSF)
of high energy $\gamma-$ray telescopes is roughly proportional to 
$E_{\gamma}^{-1/2}$. 
Observations of Galactic plane sources at energies below a few hundred MeV 
are plagued by
source confusion and low signal-to-noise 
and will continue to be so in the {\it GLAST} era. 
If the energy dependence of the PSF is included in the analysis, sources 
with significant GeV emission
can be more reliably localized and are less likely to be confused with 
background sources. A general steepening of the diffuse Galactic $\gamma-$ray 
emission (\cite{h97}) above 1 GeV further helps the signal-to-noise ratio.

In this article, we present results of X-ray and radio 
studies of the bright GeV sources. We will first discuss the lower energy
properties of the known and proposed source classes. We will then 
summarize the 
results of a recently completed 2--10 keV ASCA survey of nearly the 
entire sample 
(\cite{rrk01}, hereafter RRK) and report on analysis of VLA and ATCA radio images of
selected fields. 


\section[Observational Characteristics]
{Observational Characteristics of Potential Lower Energy Counterparts}

In general, when trying to find evidence of particle acceleration in the
Galactic plane, one looks for evidence of non-thermal emission. In X-rays,
it is necessary to observe at energies above $\sim2$ keV, both to see
through the high absorption in the  plane and to look for a hard power-law 
component of any emission. At radio wavelengths, 
one looks for a steep spectrum and high
polarization. A high radio to infrared ratio is also a good indication
of a non-thermal origin of radio emission.
Blazars, young pulsars, and SNR can all show these
characteristics allowing identification as potential $\gamma-$ray sources 
through a combination of their X-ray and radio emission.
	
Blazars are distinguished from other active galactic nuclei 
by being radio loud ($F_R \ga 0.5$
Jy). They also have moderate X-ray emission ($F_X \ga 10^{-13}$ ergs 
${\rm cm}^{-2} \, {\rm s}^{-1}$, \cite{sam97}) with hard power-law spectra 
($\Gamma \sim 1.7$).
It is therefore relatively easy to rule out a blazar identification for most
of the unidentified sources by looking at single dish radio survey data. If 
there is no bright, point-like radio source, then the source is probably
not a blazar. Blazars are quite variable at most wavelengths, including
$\gamma-$rays, unlike the known pulsars and what is expected from SNR shells.
Therefore, any variable $\gamma-$ray source that is not a blazar is
likely to belong to a new class of objects.

\begin{figure}[ht]
\centerline{\epsfig{file=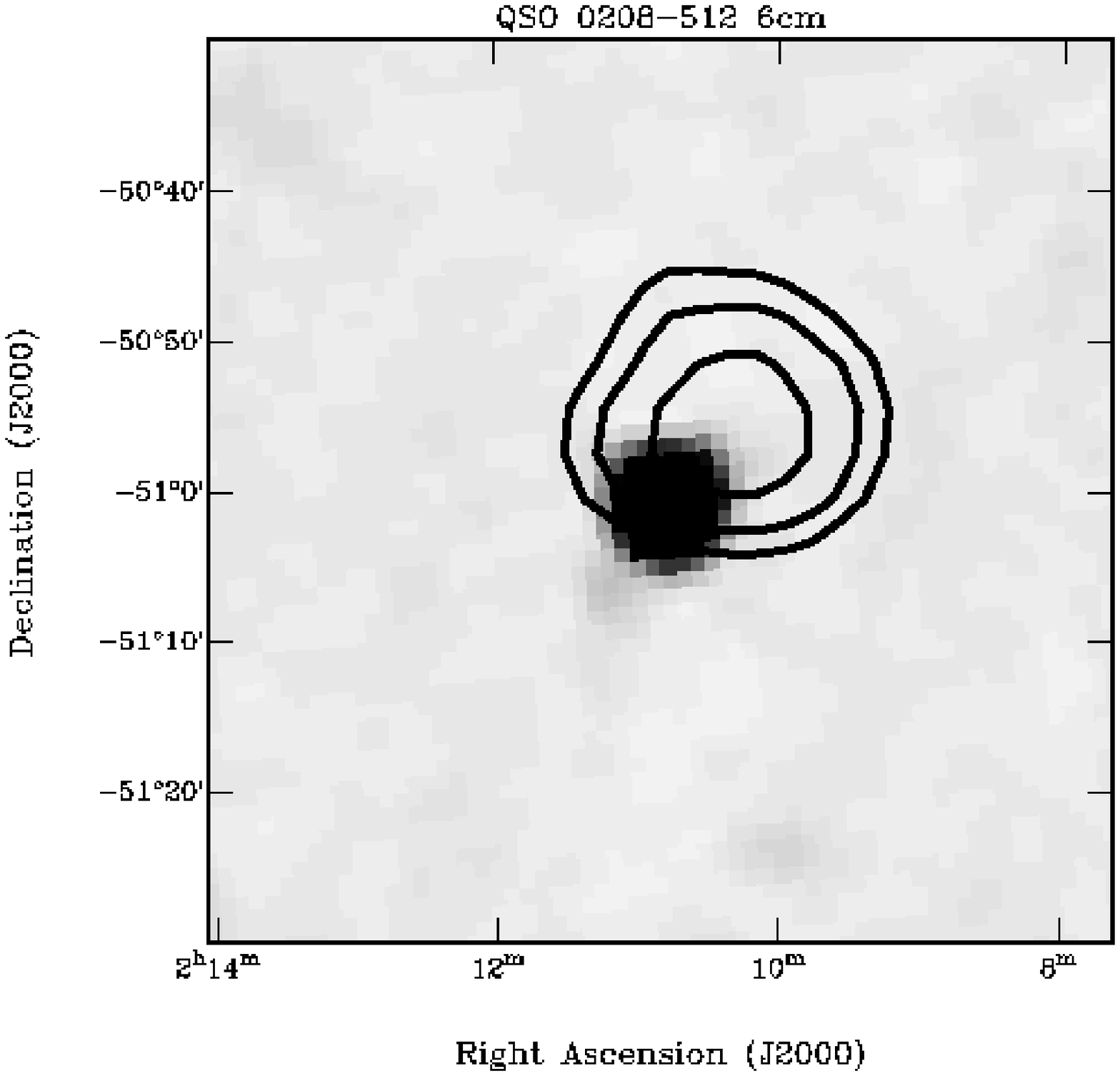, width=0.4\hsize}
            \epsfig{file=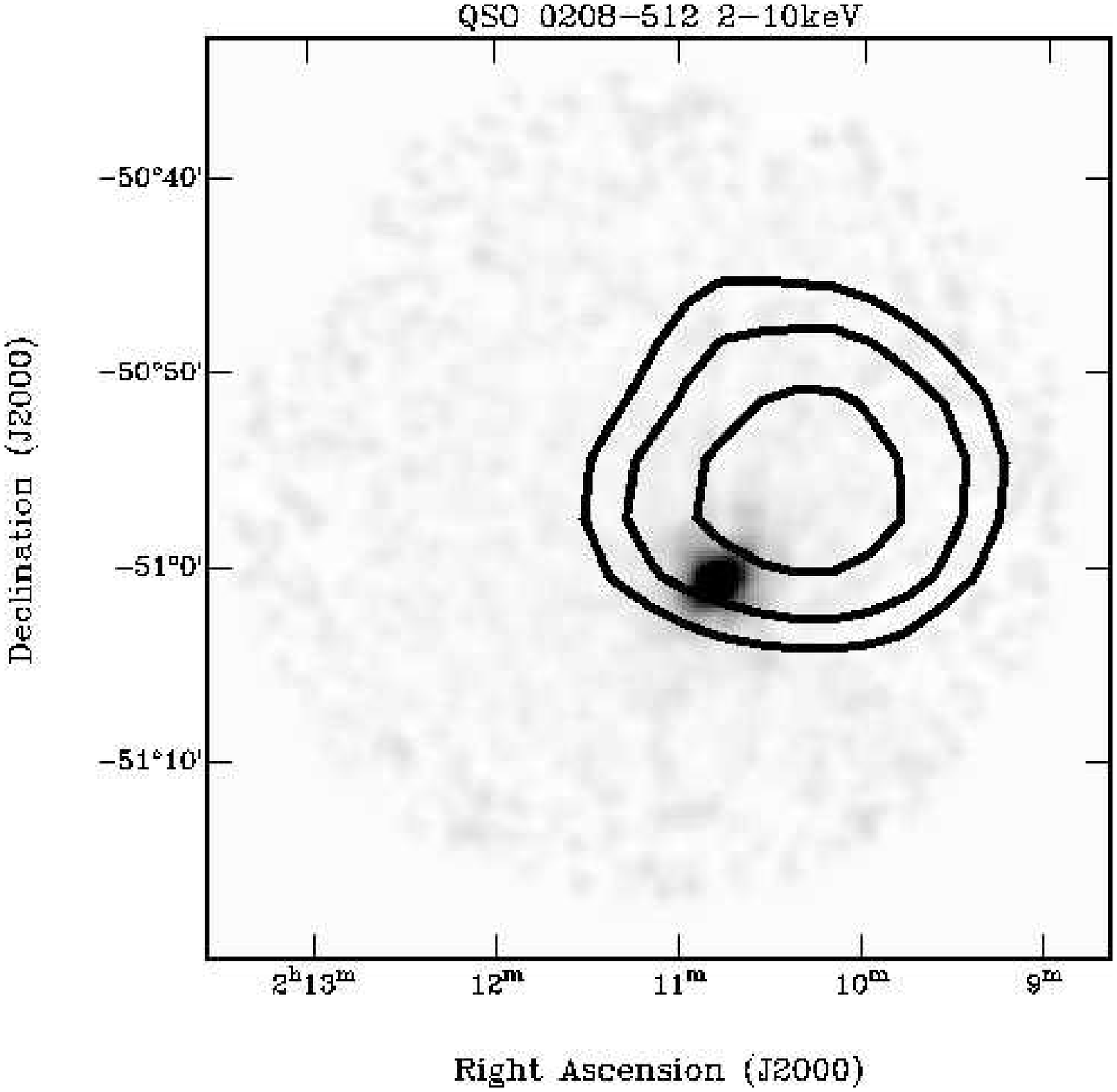,    width=0.4\hsize}}
\caption{{\bf Left:} 4850 MHz radio image of the GeV blazar 
QSO 0208$-$512 from the PMN survey (\cite{c91}). The contours mark the 
 $EGRET$ 68\%, 95\%, and 99\% confidence position regions. {\bf Right:}
2-10 keV ASCA image of QSO 0208$-$512.}
\end{figure}

The known $\gamma-$ray pulsars can be placed in two rough
categories. The youngest ones (Crab and Vela)  have bright synchrotron 
X-ray and radio 
wind nebulae surrounding them which dominate
the emission at lower energies.
These pulsar wind nebulae (PWN) are a common feature of the youngest pulsars
($\tau \la 10^4$ years old). Older $\gamma-$ray pulsars, 
such as PSR 1055$-$52 and
Geminga, appear as point sources in X-rays with a spectrum consisting
of a soft thermal component (often highly absorbed) and a flat
power law component (photon spectral index $\Gamma \la 2$, \cite{hw97}). These have
weak or no PWN, and will be referred
to here as isolated pulsars. The known $\gamma-$ray pulsars have not been 
observed to be variable in studies of $E \ge 100$ MeV {\it EGRET} emission
(\cite{m96,t99}). However, the Crab PWN appears variable to {\it EGRET} in the 
soft 70-150 MeV energy band (\cite{dj96}).

Supernova remnants are identified at radio wavelengths 
by having a shell
type morphology, a steep radio spectrum, and significant polarization. 
They can also be distinguished from thermal radio sources, such
as HII regions, by a high radio to infrared emission ratio
(\cite{wg96}). PWN
are distinguished
in radio from SNR by having somewhat flatter spectra 
(energy spectral index $\alpha \sim -0.3 - -0.1$), somewhat higher 
polarization fraction ($\sim 10 - 30\%$), and amorphous morphologies
(\cite{fs97}). In X-rays, SNR can have either a shell or filled center
morphology with thermal plasma spectra having strong emission lines. 
PWN tend to have featureless, hard power-law X-ray spectra with toroidal, 
bow-shaped, or jet-like morphologies. Since SNR are generally several 
parsecs across, they are not expected to vary on timescales observable by 
{\it EGRET}, unless
the $\gamma-$ray emission is highly localized.



\section{ASCA X-Ray Survey}

\begin{figure}[h!]
\centerline{\epsfig{file=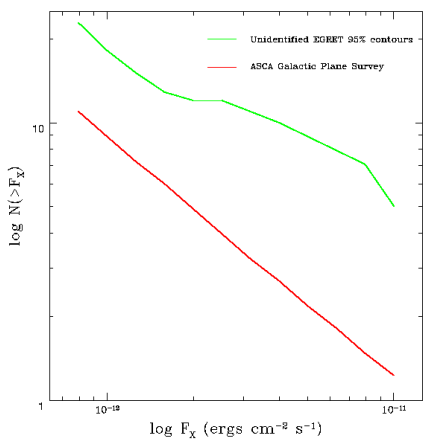, width=0.5\hsize}}
\caption{The $ASCA$ 2--10 keV log N--log S distribution of 
X-ray sources found in the 95\% error contours of unidentified GeV 
sources (RRK) compared to the $ASCA$ Galactic Plane Survey (\cite{s01}).
\label{lnls}}
\end{figure}

The images in this paper (unless otherwise noted) were made with the GIS 
instrument on the {\it ASCA} satellite. They come from the 
{\it ASCA} catalog of GeV sources 
(RRK) which covers about 85\% of the
{\it EGRET} 95\% error contours for 19 of the 20 bright unidentified 
GeV sources listed in LM97. 
In order to screen out stellar sources, which tend
to be quite bright in soft X-rays but faint above 2 keV, 
2--10 keV images that have been
exposure corrected and smoothed are shown. Some images are composites
from multiple pointings, and sometimes field edge artifacts appear 
from scattered light. The details of the image reduction and analysis 
can be found in RRK.
A log N--log S comparison in the 2--10 keV band of the 
unidentified GeV source fields to the relation derived 
from the {\it ASCA} Galactic Plane Survey 
($|l| < 45^{\circ} ,|b| < 0.4^{\circ} $, \cite{s01})
shows a clear excess of sources with $F_{2-10\rm keV}\ga 10^{-12}$ ergs 
${\rm cm}^{-2} {\rm s}^{-1}$ (Figure~\ref{lnls}).
    
\subsection{Previously Known Supernova Remnants}

\begin{figure}[h!]
\centerline{\epsfig{file=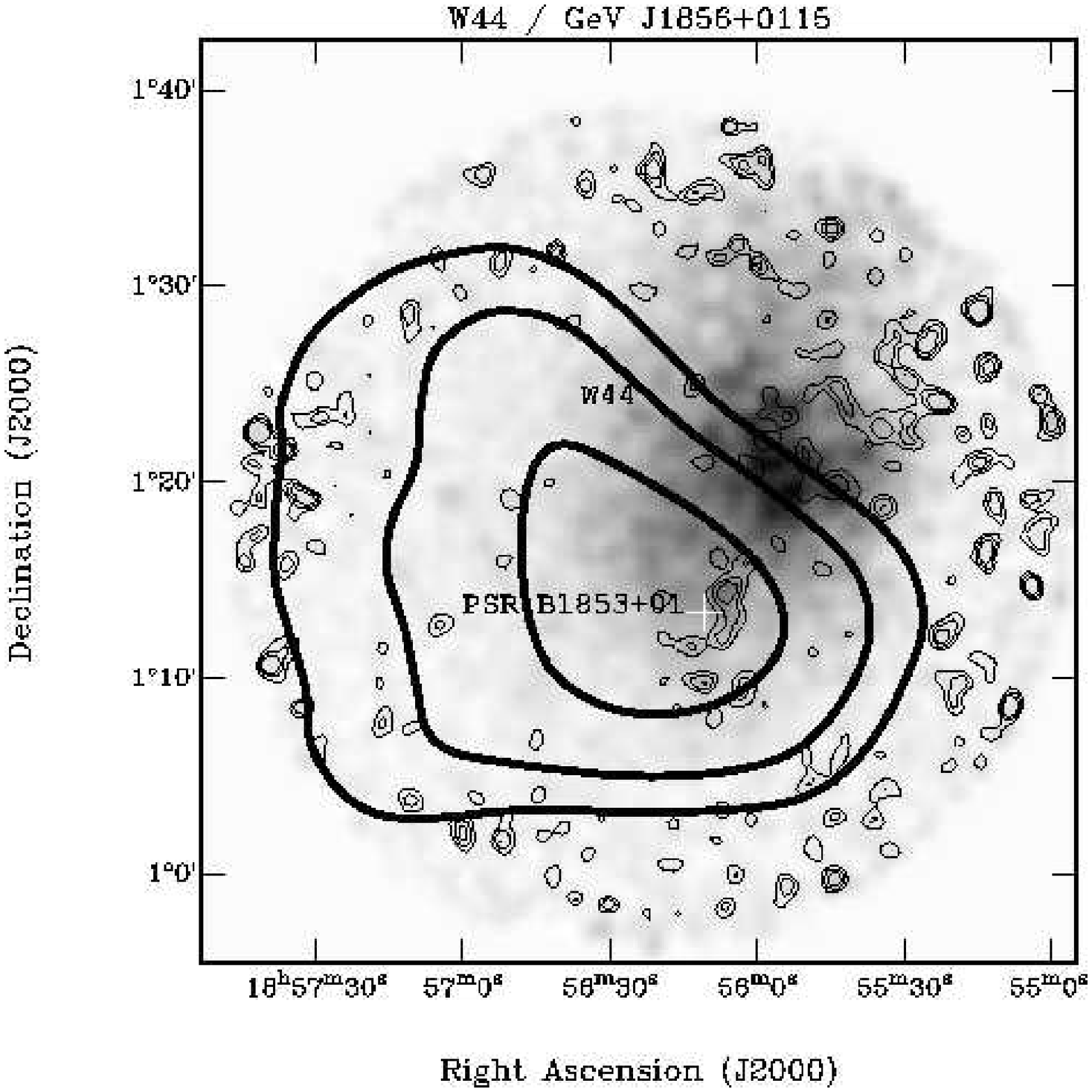, width=0.4\hsize}
            \epsfig{file=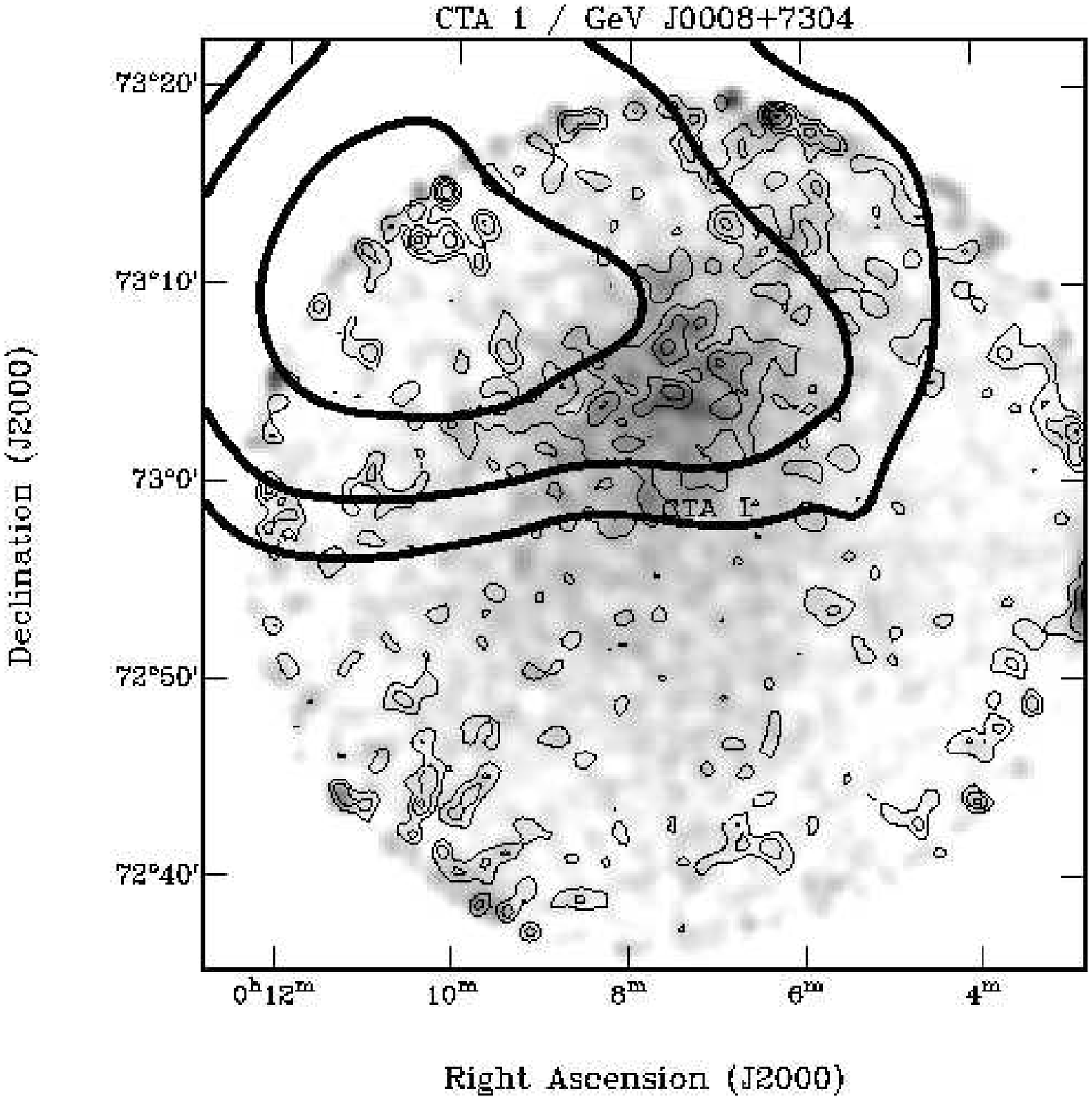,    width=0.4\hsize}}
\caption{{\bf Left:} The SNR W44 in 2-10 keV X-rays, with 4-10 keV
light contours. The heavy contours are the $EGRET$ 68\%, 95\%, and 99\%
confidence positions of GeV J1856+0114. {\bf Right:}
SNR CTA 1 X-ray image with hard X-ray and GeV J0008+7304 contours. \label{w44}}
\end{figure}

There are five well observed SNR coincident with bright unidentified
GeV sources. The X-ray images tend to be dominated by thermal plasma emission
which falls off rapidly above 2 keV. Therefore, 4--10 keV
contours on the 2--10 keV image can reveal a 
hard, power-law spectral component which is evidence of
particle acceleration. W44 (GeV J1856+0115) and CTA 1 (GeV J0008+7304) 
both contain hard emission coincident
with the GeV source (Figure~\ref{w44}). However, the hard source in W44 is a PWN associated
with the young pulsar PSR B1853+01 (\cite{hhh96}), while the emission in 
CTA 1 is a suspected PWN associated with the X-ray pulsar candidate 
RXJ0007.0+7302 (\cite{s97}). This suggests the GeV emission is
related to the pulsars and not the SNR shell.

\begin{figure}[h!]
\centerline{\epsfig{file=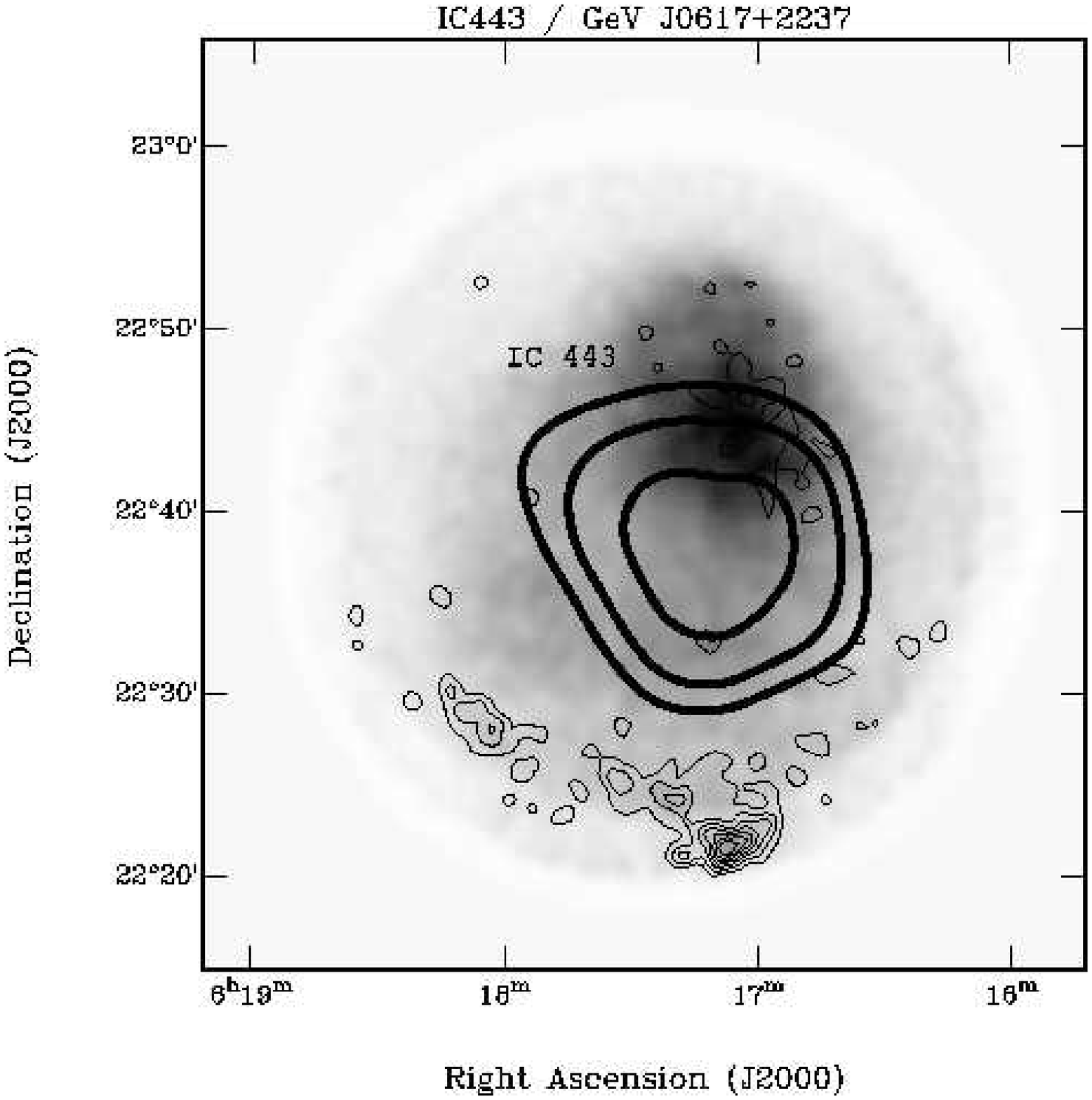, width=0.4\hsize}
            \epsfig{file=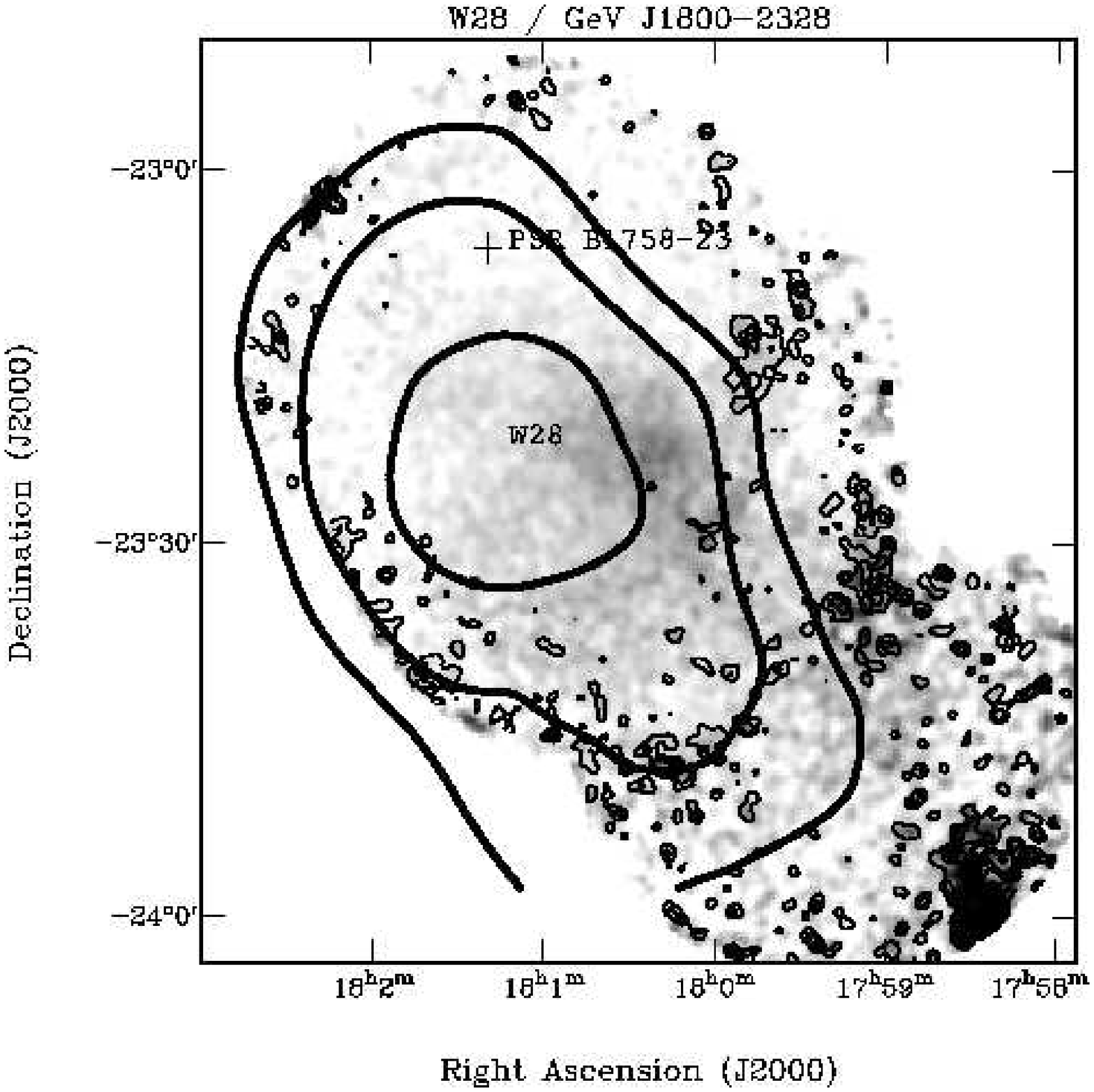,    width=0.4\hsize}}
\caption{{\bf Left:} SNR IC443 / GeV J0617+2237. {\bf Right:}
SNR W28 / GeV J1800$-$2328. \label{w28}}
\end{figure}

IC443 (GeV J0617+2237) and W28 (GeV J1800$-$2328) also have regions of 
hard X-ray emission, but in these cases
the emission is not well correlated with the {\it EGRET} positions derived from
GeV photons
(Figure~\ref{w28}).  The two regions of hard X-ray emission in the
southern portion of the IC443 {\it ASCA} field, 
one of which now appears to be a 
pulsar with PWN (\cite{k01}), have previously been suggested
as sites of shock acceleration producing GeV emission (\cite{k97}).
The newer GeV based position seems to exclude this. W28 is coincident
with the young pulsar PSR B1758$-$23. However, the high dispersion 
measure distance of 13.5 kpc (\cite{k93}), although admittedly
unreliable, suggests the pulsar is not associated with the remnant or 
GeV J1800$-$2328.

GeV J2020+4023 is well located near the center of the $\gamma$ Cygni
SNR shell. Despite very extensive observations with ASCA, there
is no obvious X-ray emission associated with the $\gamma-$ray source.
Therefore, none of the SNR associated with GeV sources show evidence of 
X-ray shock emission consistent with the GeV position which can not be
attributed to a young pulsar. 

\begin{figure}[h!]
\centerline{\epsfig{file=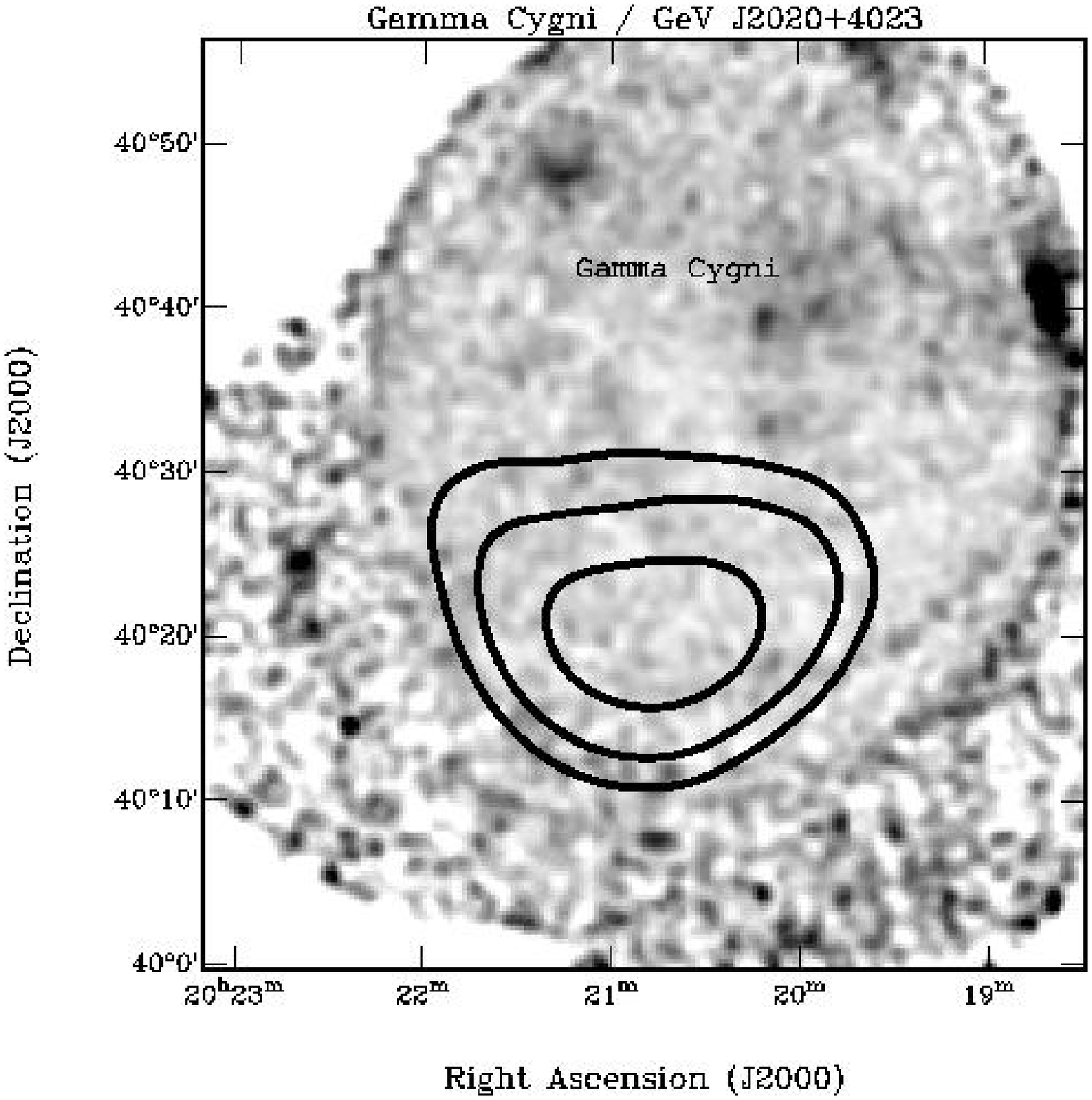, width=0.4\hsize}
            \epsfig{file=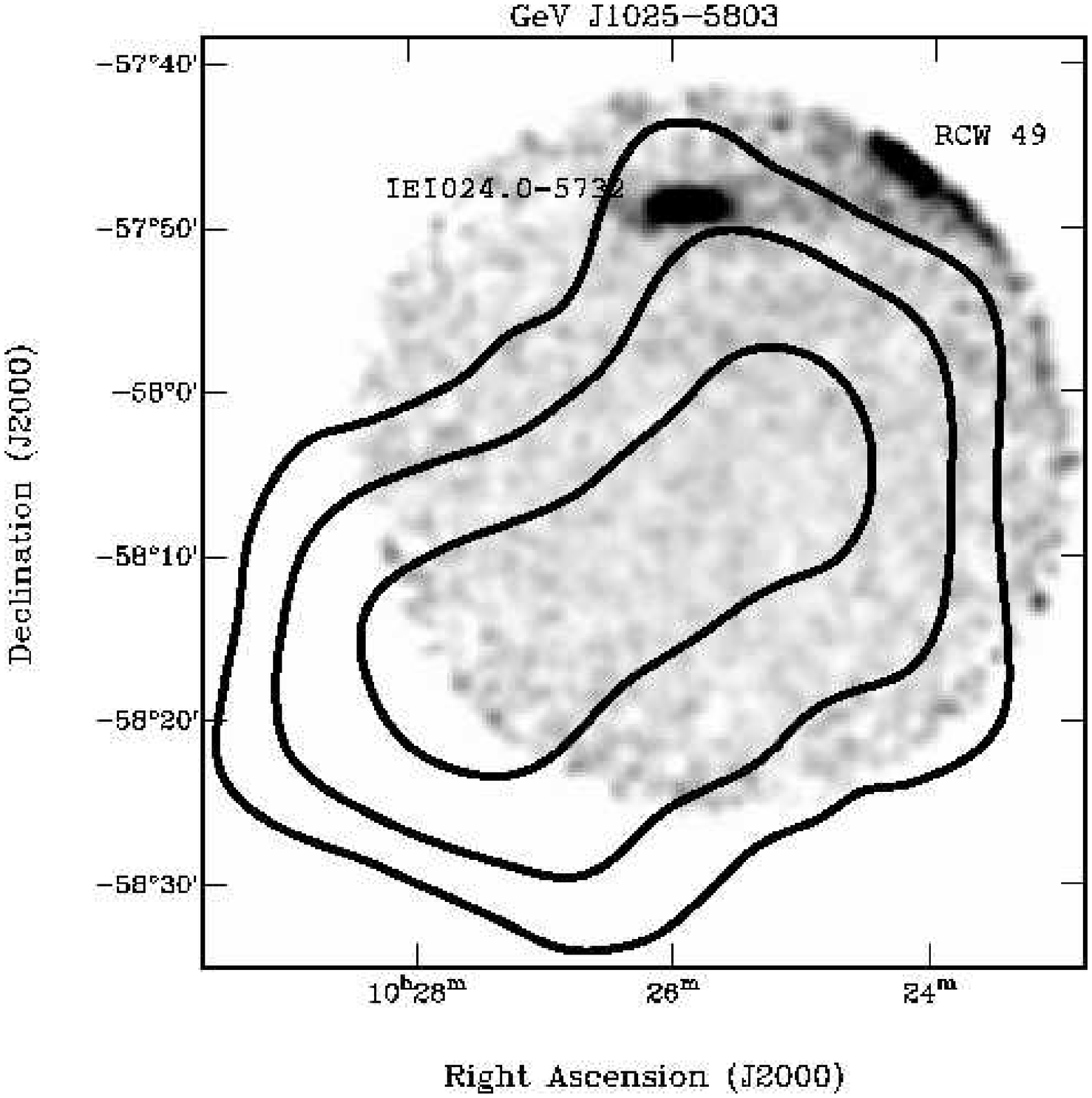,    width=0.4\hsize}}
\caption{{\bf Left:} 2-10 keV image of SNR $\gamma-$Cygni / GeV J2020+4023. {\bf Right:}
GeV J1025$-$5803 with the WN5+0 binary star 1E1024.0$-$5742. At the edge
of the image is a colliding wind system in the RCW49 HII region. 
\label{g1025}}
\end{figure}

\subsection{Massive Binaries}

\begin{figure}[h!]
\centerline{\epsfig{file=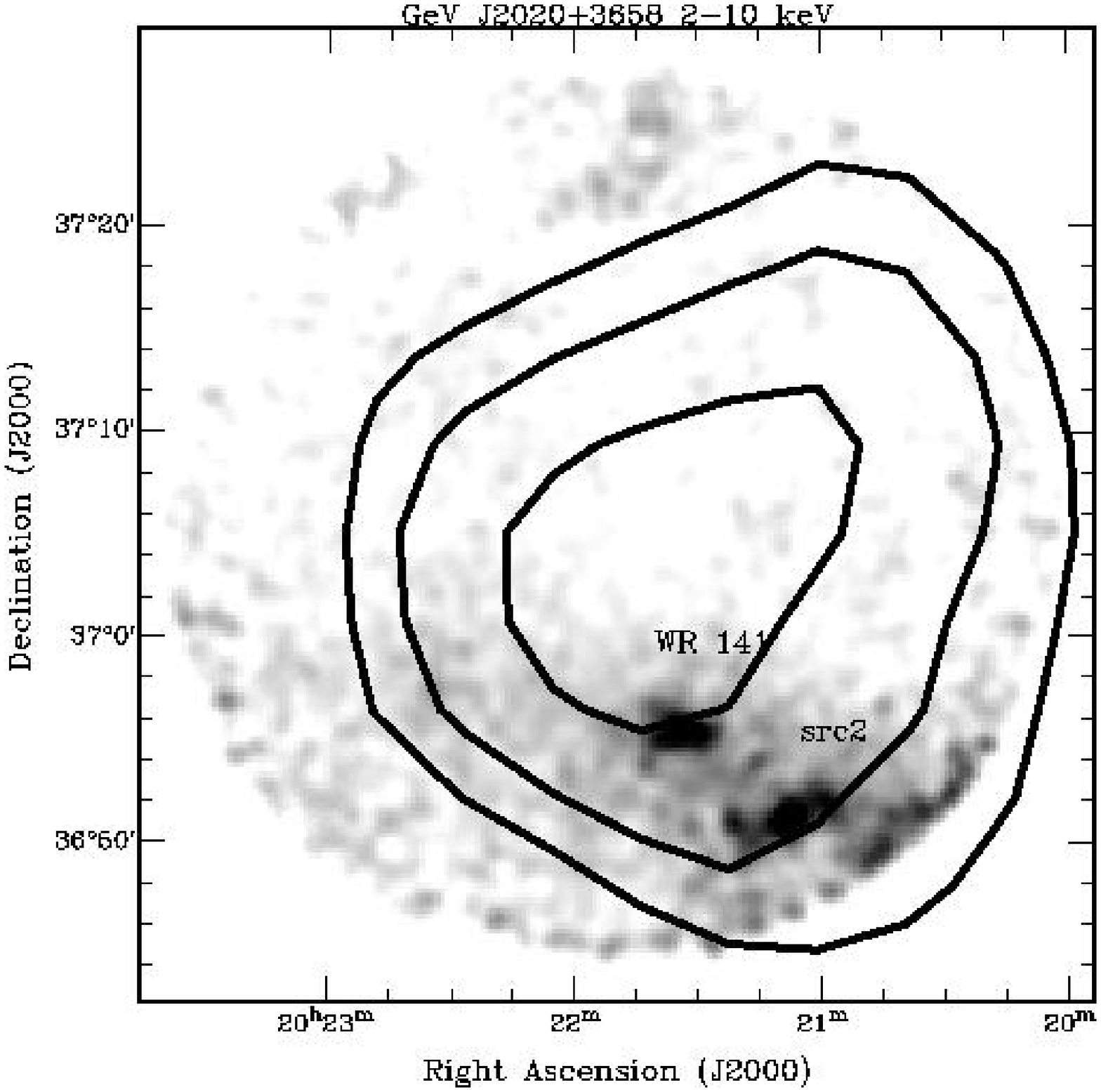, width=0.4\hsize}
            \epsfig{file=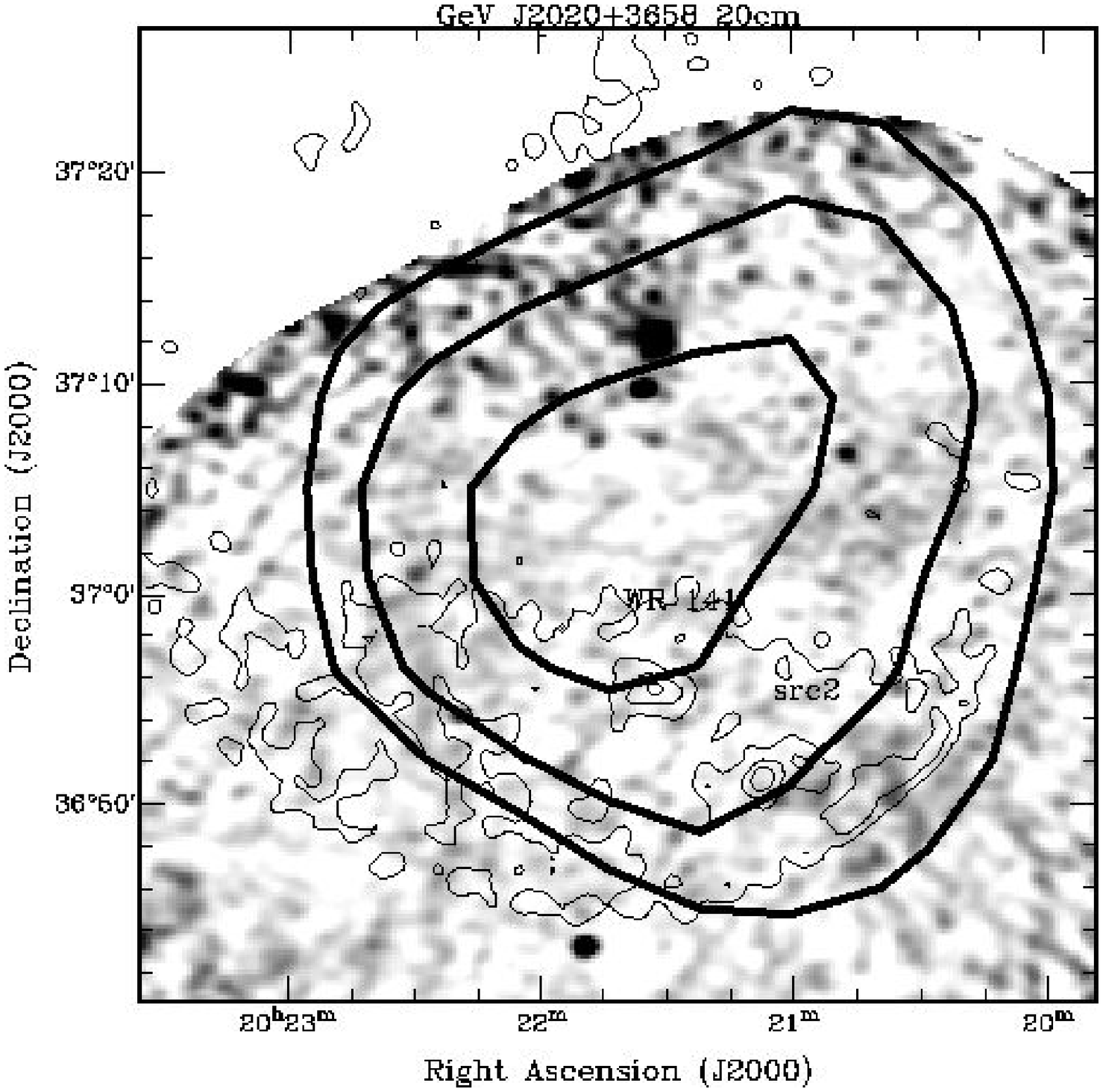,    width=0.4\hsize}}
\caption{{\bf Left:} GeV J2020+3658 in X-rays, showing WR 141
and the second hard source embedded in diffuse X-rays. {\bf Right:}
20 cm VLA image of GeV J2020+3658 field with X-ray as light contours. \label{g2019}}
\end{figure}

Colliding winds in massive binary systems are potential sources of
$\gamma-$ray emission (\cite{mp01}), and it is interesting to note any candidates among
the bright GeV sources. There are two Wolf-Rayet + O star binary systems:
1E1024.0$-$5742 consistent with GeV J1025$-$5809 and WR141 consistent with
GeV J2020+3658. Neither can be considered a 
very strong candidate. 1E1024.0$-$5742
is outside the somewhat large 95\% positional contour. WR141 is well within the
positional contour, but so is a second, equally bright hard point source.
These two sources in GeV J2020+3658 are embedded in diffuse X-ray emission 
with a somewhat steep (photon index $\Gamma \sim 2.5$) spectrum, suggestive
of a thermal SNR. However, a 20~cm observation of the field with the 
VLA in D array shows no bright radio nebula (Figure~\ref{g2019}). 

GeV J0241+6102 is consistent with the peculiar radio emitting
Be-star/X-ray binary LSI+61 303. This source has been the subject of much
study (cf. \cite{st98}), but the nature of the compact object in the
system is still uncertain.
Although the $\gamma-$ray source is moderately
variable, the variations have not been correlated with any known
time scales of LSI+61 303. With the latest {\it EGRET} positional
determination, the source now lies just outside the
95\% confidence contour.
 
\section{Multi-Wavelength Studies of GeV Selected Fields}

\begin{figure}[ht]
\centerline{\epsfig{file=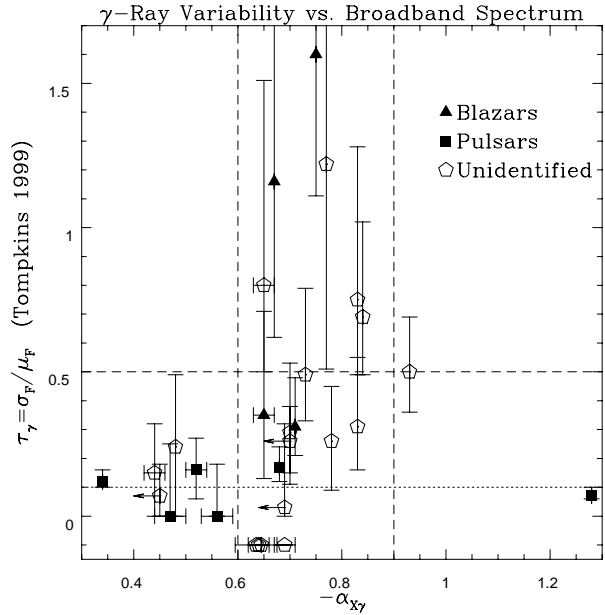, width=0.7\hsize}}
\caption{$E\ge 100$ MeV variability index $\tau\equiv \sigma_F/F$ 
versus the X-ray to $\gamma-$ray ``spectral index" $-\alpha_{X\gamma}$
assuming the brightest non-thermal X-ray source consistent with the GeV source
as the counterpart. 
Dashed lines denote regions of high and low variability, and
from left to right X-ray faint, moderate, and bright. The dotted line is the level
of variability due to systematic errors.
\label{atplot}}
\end{figure}

A crucial component in further categorizing the sources is 
$\gamma-$ray variability. Tompkins (1999) has measured the variability
of all the sources listed in the third {\it EGRET} catalog (\cite{h99})
using the $\tau\equiv \sigma_F/\mu_F$ statistic, where $\sigma_F$ is
the standard deviation of the measured fluxes from different viewing
periods 
and $\mu_F$ is the mean flux from all viewing periods. Note this is a measure
of how variable a source is rather than the usual $\chi^2$ test of how 
inconsistent the source is with a 
constant model (cf. \cite{m96}). In Figure~\ref{atplot} 
we plot all 28 sources in RRK the $\tau$ value versus 
$-\alpha_{X\gamma}\equiv -(1+\log(F_{\gamma}/A_X)/6)$, a broadband
energy 
`spectral index' where $F_{\gamma}$ 
is the photon 
flux above 1~GeV (equivalent to the power-law normalization at 
1~GeV for a source with photon index $\Gamma=2$) and $A_X$ is the 
1~keV X-ray power law normalization in 
photons ${\rm GeV}^{-1}\,{\rm cm}^{-2}$. 
In fields with more than one X-ray source, 
the $-\alpha_{X\gamma}$ values are derived from the brightest
source consistent with being the GeV source, and can therefore be 
considered an upper limit. The plot is split 
into regions of high and low variability, and, from left to right, X-ray
faint, moderate, and bright. The dotted line represents the
systematic uncertainty in the $\tau$ measurement, which is consistent
with no variability. 

The only X-ray bright sources are the
Crab nebula and possibly the source near the Galactic center. 
Vela is the only X-ray moderate pulsar, due to its PWN. The 
isolated pulsars are X-ray faint. All the identified GeV 
pulsars are consistent with no variability ($\tau=0.1$,
the systematic variability of {\it EGRET}), and the
blazars are significantly variable. Four GeV
sources in this sample are not in the 3EG catalog, and so were not 
measured for variability by Tompkins. These are shown along the bottom.

\subsection{Isolated Pulsar Candidates}   

There are three unidentified sources in the X-ray faint, low variability
category which also contains the isolated pulsars. 
One is the source in SNR $\gamma-$Cygni. Since there are
only weak upper limits on the non-thermal emission
in W28 and IC 443, the $\gamma-$ray sources positionally coincident 
with these
SNR may also fall in this category.
The other two sources are GeV J1835+5921 and GeV J1837$-$0610. 
GeV J1835+5921 is the only high latitude source among the
bright unidentified GeV sources, and has been extensively studied
by Mirabal et al. (2000). Through X-ray, optical and radio 
studies, they identify every source in the field and find a single faint 
X-ray source whose ratio of X-ray to optical luminosity is 
consistent with a neutron star identification.
  
\begin{figure}[h!]
\centerline{\epsfig{file=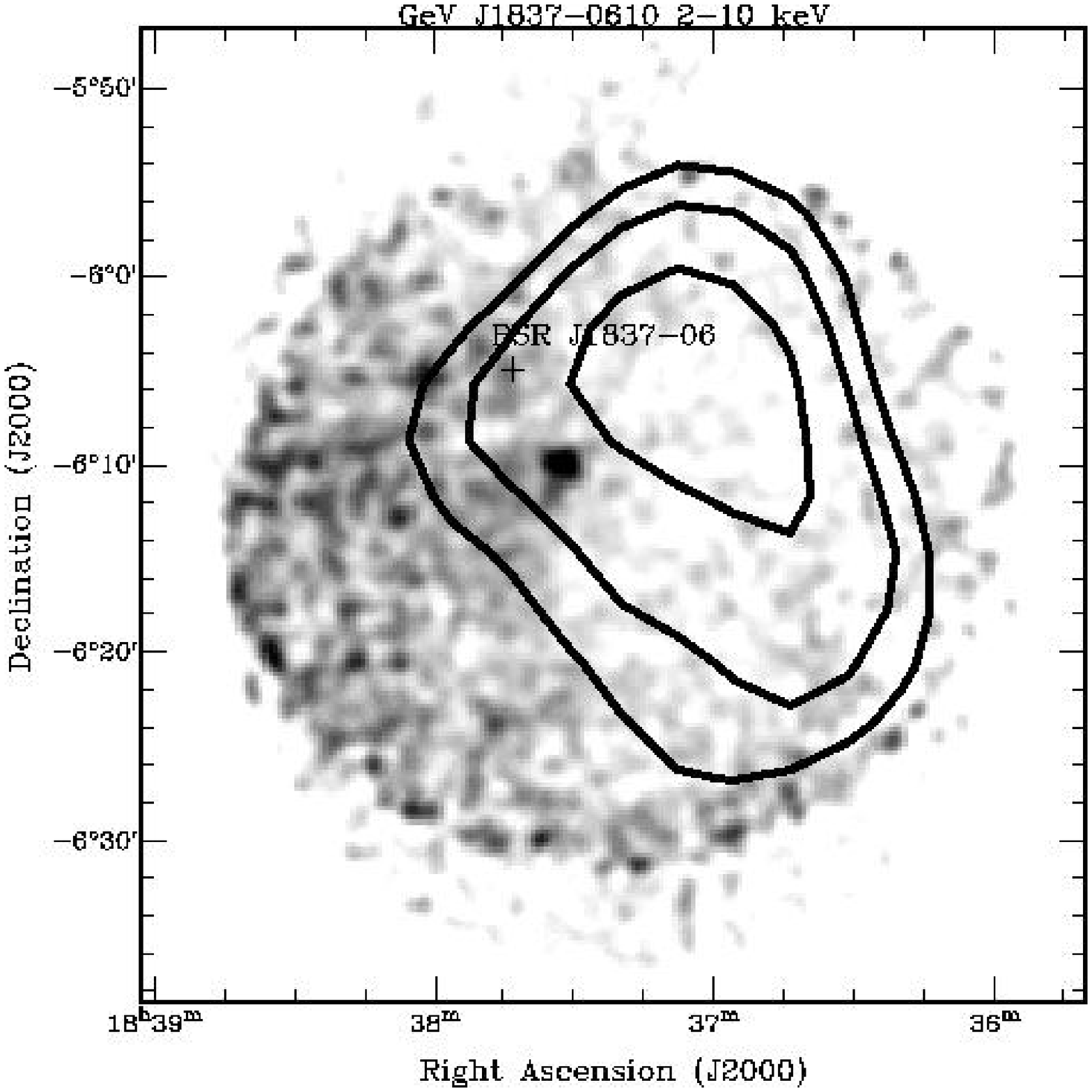, width=0.4\hsize}
            \epsfig{file=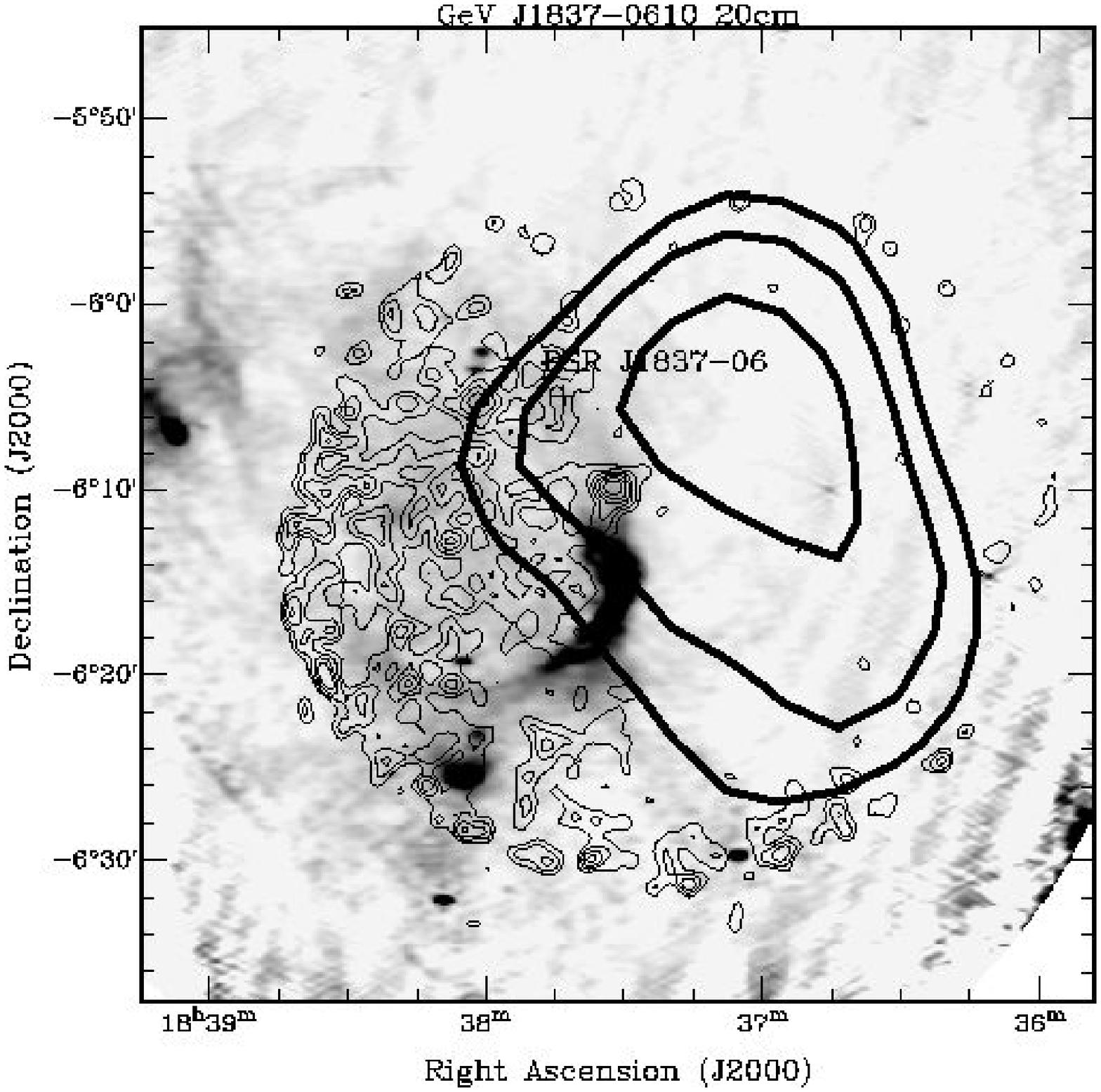,    width=0.4\hsize}}
\caption{{\bf Left:} GeV J1837$-$0610 in 2-10~keV X-rays, with the
position of the young pulsar marked. {\bf Right:}
20 cm VLA image of GeV J1837$-$0610 field with X-ray contours. \label{g1837}}
\end{figure}

The field of GeV J1837$-$0610 contains one hard X-ray point source 
in the {\it ASCA} image (Figure~\ref{g1837}) and faint diffuse emission. A 20 cm VLA
D array image shows a faint ring surrounding the
X-ray emission with a very bright HII region
along one edge. This is suggestive of an old SNR interacting with the 
molecular cloud. However, the dynamic range requirements caused by
the HII region make accurate spectral measurements of the ring
difficult. Also in the field is the young (characteristic age
of 34,000 yr) pulsar PSR J1837$-$0604 recently
discovered by the Parkes Multibeam Survey (\cite{d01}). 
The dispersion measure distance to the pulsar of 6.2 kpc is consistent 
with the distance to the molecular cloud.

\begin{figure}[h!]
\centerline{\epsfig{file=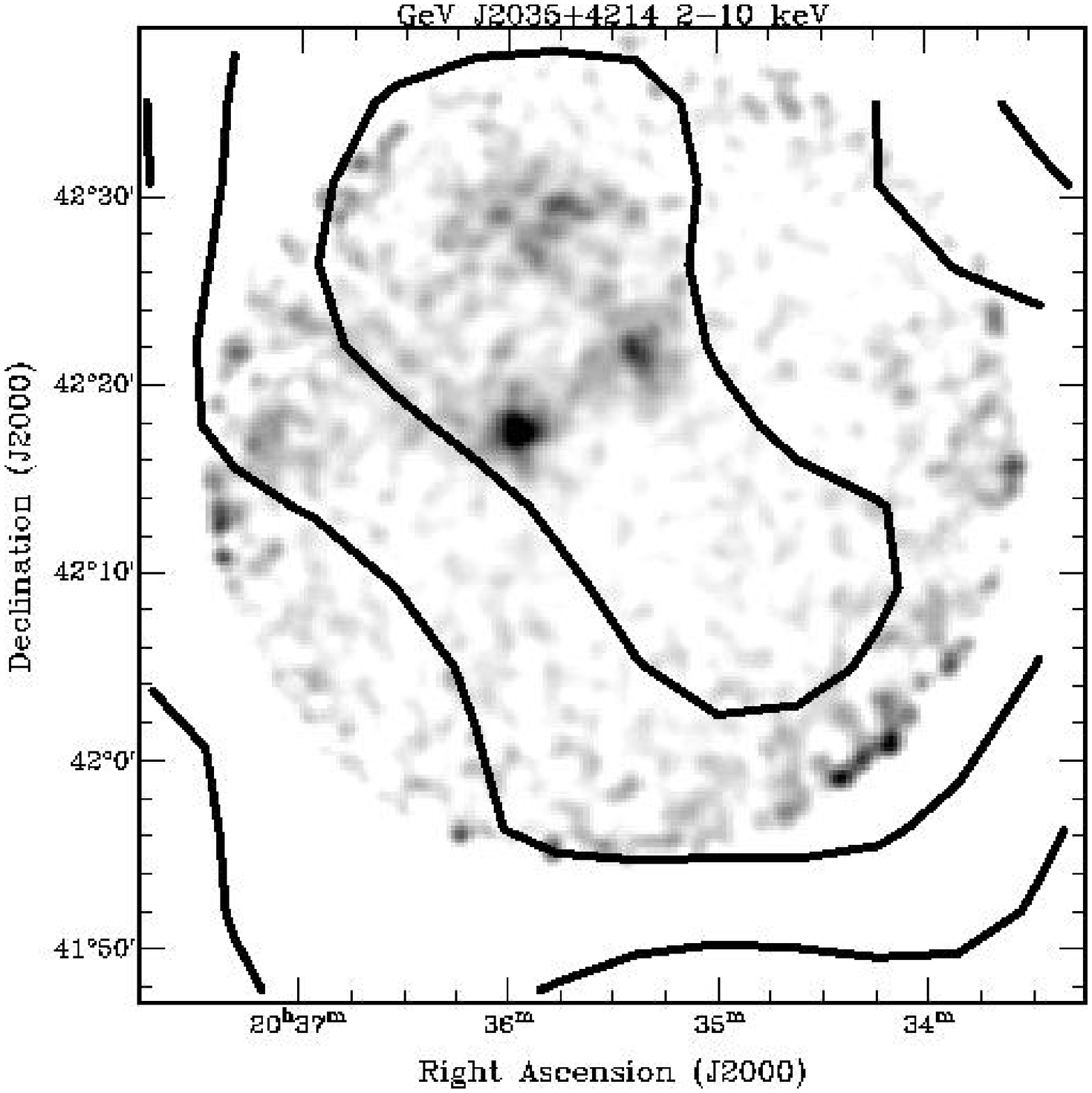, width=0.4\hsize}
            \epsfig{file=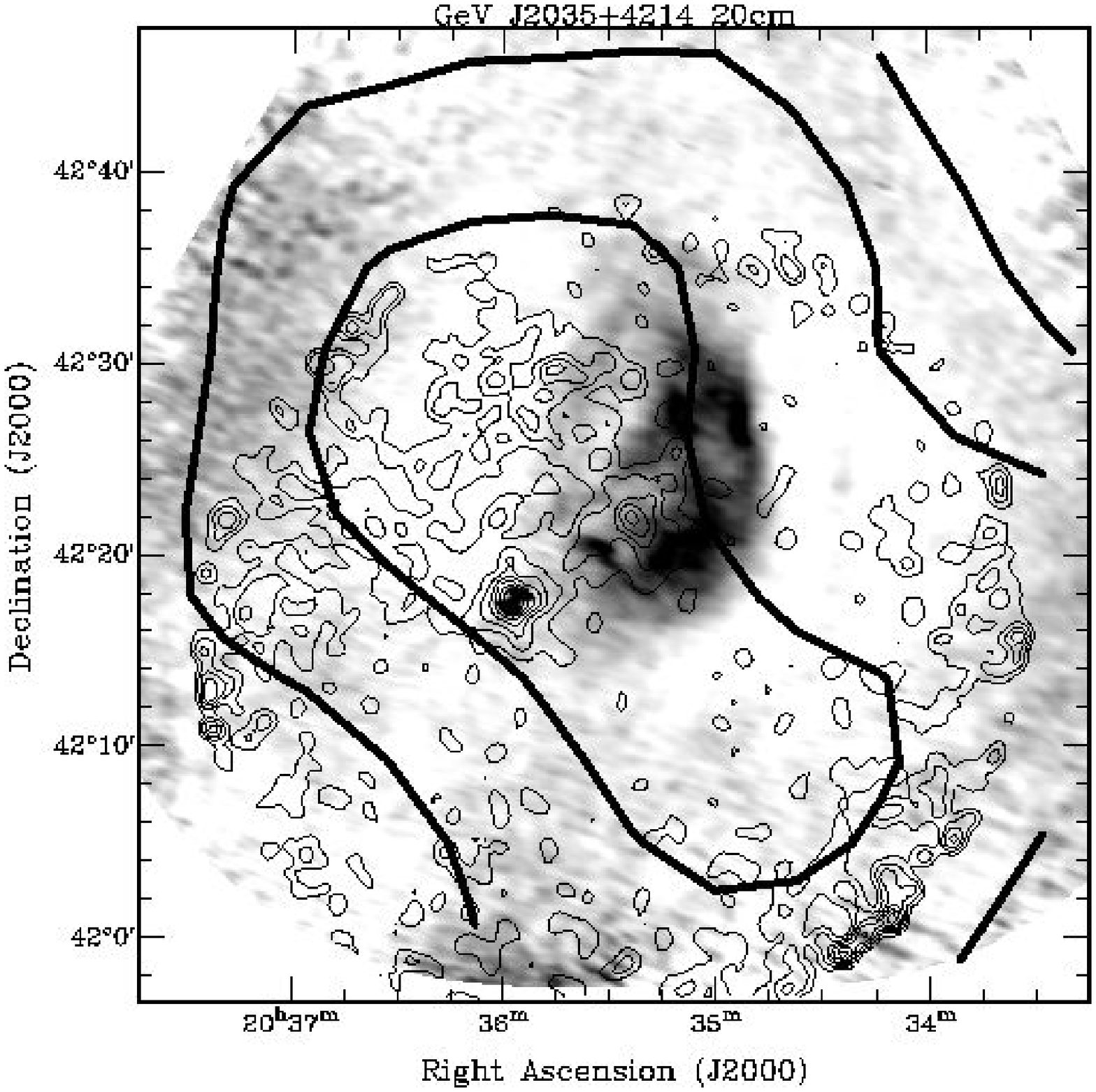,    width=0.4\hsize}}
\caption{{\bf Left:} GeV J2035+4214 in X-rays. {\bf Right:}
20 cm VLA image of GeV J2035+4214 field showing the bright
molecular cloud DR17. \label{g2035}}
\end{figure}

GeV J2035+4214 is in the X-ray moderate category and was not in
the 3EG catalog and so was not measured by Tompkins for variability. 
The X-ray image (Figure~\ref{g2035}) shows three sources: two hard point sources and 
a softer diffuse source. In the VLA 20 cm image it can be seen that one of 
the point sources is associated with a double-lobed radio galaxy. The
other is embedded in the bright, nearby ($\la 1$ kpc) molecular cloud
DR17. The intensity of the brightest features of the molecular cloud 
make it difficult to determine if the dim region near the X-ray source 
has a different spectrum from the rest of the cloud. A somewhat speculative
possibility for the emission from GeV J2035+4214 and GeV J1837$-$0610 is
that shock accelerated particles from an old SNR (whose diffuse thermal 
X-ray emission we still see)
are interacting with the molecular clouds, causing the GeV emission.
However, the interpretation as isolated pulsars may be more conservative. 

\subsection{Candidate Pulsar Wind Nebulae}

\begin{figure}[h!]
\centerline{\epsfig{file=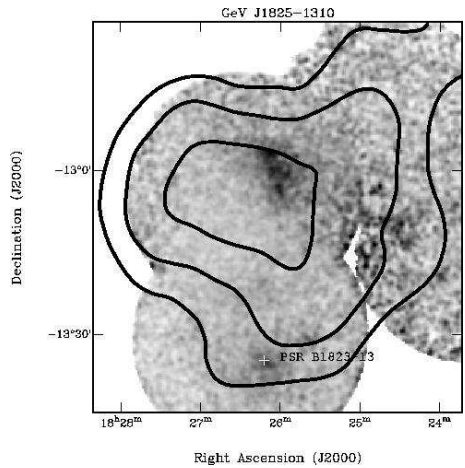, width=0.5\hsize}}
\caption{GeV J1825$-$1310 at 2-10 keV. The X-ray nebula 
is well centered with the young pulsar PSR B1823$-$13 to the south.
The excess emission near where the four pointings join is from G18.1-0.2,
a probable SNR in the Sharpless 53 HII cluster (\cite{k89}), which is also 
consistent in position with the nearby softer $\gamma-$ray source
3EG J1823$-$1314.  \label{g1825}}
\end{figure}

The most intriguing sources are the X-ray moderate, high variability sources.
There are four in this category, not including blazars. In addition,
the source containing LSI+61 303 is on the high variability border, and 
the source inside CTA 1 shows evidence of variability. 
One of the four high variability sources is the PWN in W44 associated with 
PSR B1853+01. A second, GeV J1825$-$1310, 
is near the young pulsar PSR B1823$-$13. However, the {\it ASCA} image of
the field (Figure~\ref{g1825}) reveals a previously unknown, hard spectrum 
($\Gamma \sim 2.2$) X-ray nebula. The other two variable
sources also contain X-ray nebula, and are discussed in detail in the
following sections.

\subsubsection{The Kookaburra and the Rabbit}

\begin{figure}[h!]
\centerline{\epsfig{file=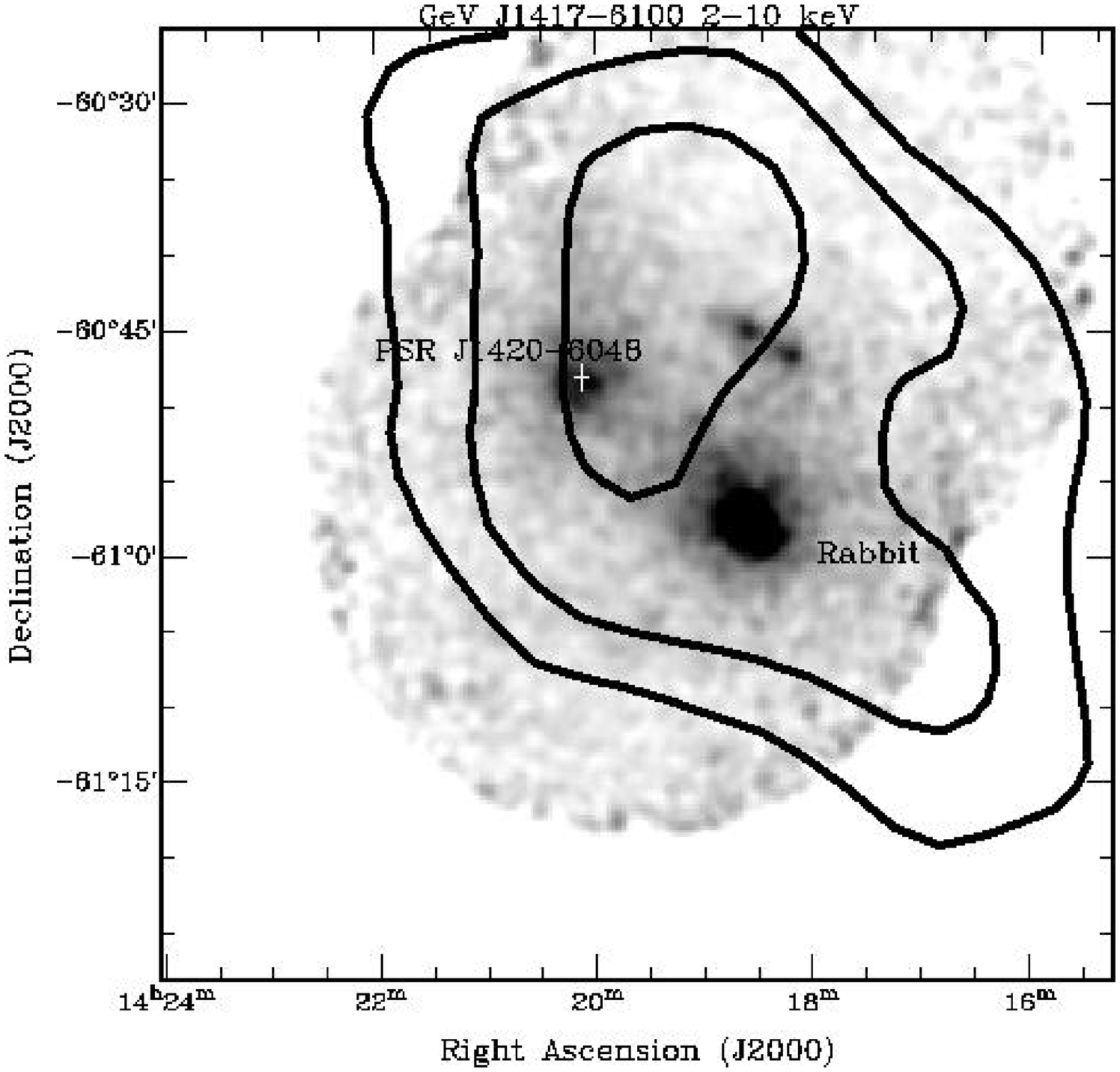, width=0.4\hsize}
            \epsfig{file=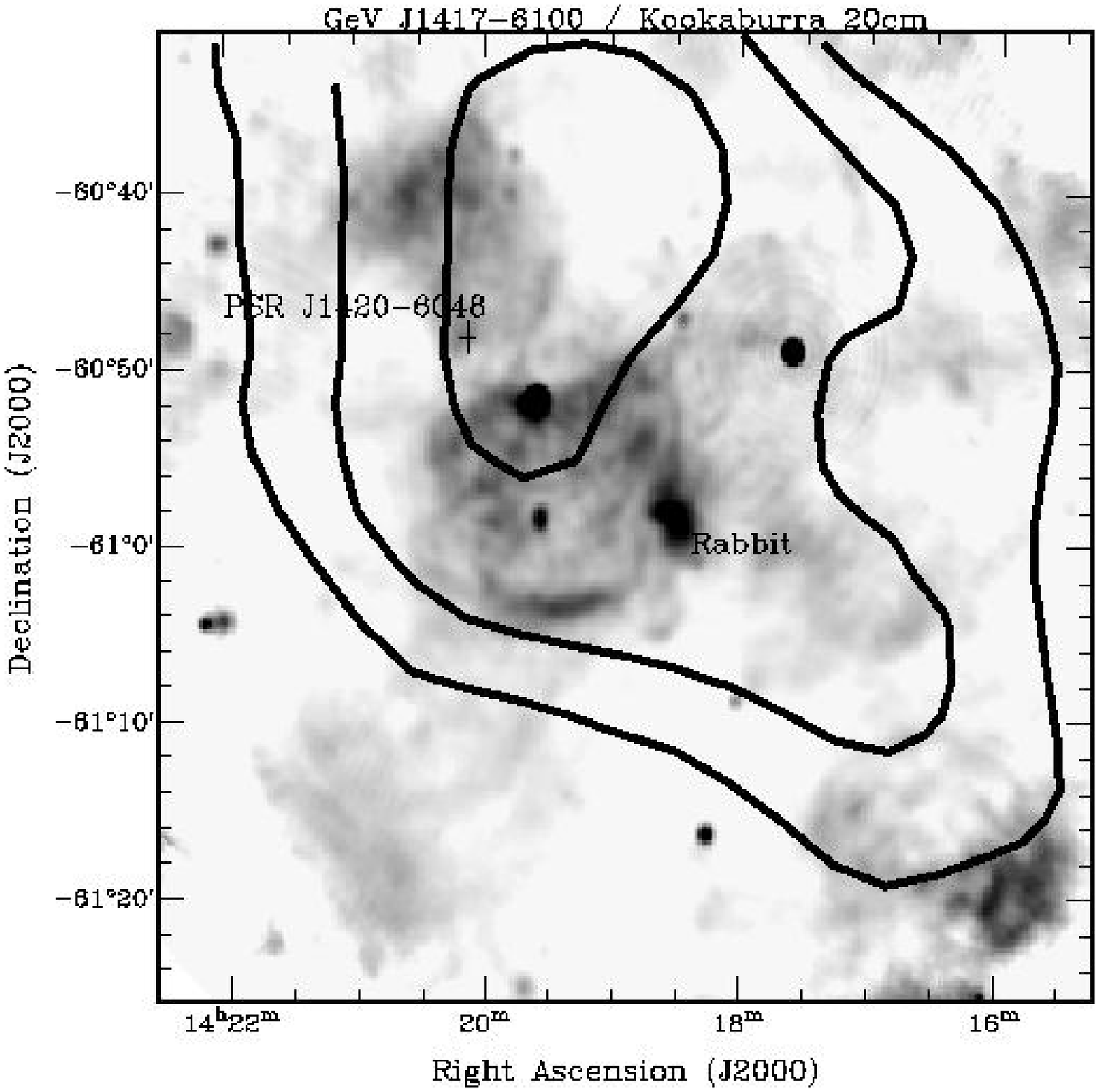,    width=0.4\hsize}}
\caption{{\bf Left:} GeV J1417$-$6100 in X-rays. {\bf Right:}
20 cm ATCA image of GeV J1417$-$6100 field showing the 
Kookaburra Nebula and the locations of the Rabbit Nebula and 
PSR J1420-6049. \label{g1417}}
\end{figure}

GeV J1417$-$6100 is another X-ray moderate high variability source
that has been the subject of much recent study (\cite{rr98,r99,cb99,r00,d01}).
The region contains the Kookaburra radio complex, within which are two
extended hard X-ray sources and two hard point sources (Figure~\ref{g1417}). One of the
point sources is coincident with a weak, variable radio source, and is 
likely to be a Seyfert galaxy. The other point source appears variable, and
may also be a Seyfert, although there is no radio counterpart. 
The Kookaburra complex itself consists of a mostly thermal shell, with
two wings extending to the north and south which, from comparison to
infrared images and some excess polarization, may be non-thermal. 
At the edge of the shell, near the southern wing, is a moderately bright 
radio nebula known as the Rabbit. 

\begin{figure}[h!]
\centerline{\epsfig{file=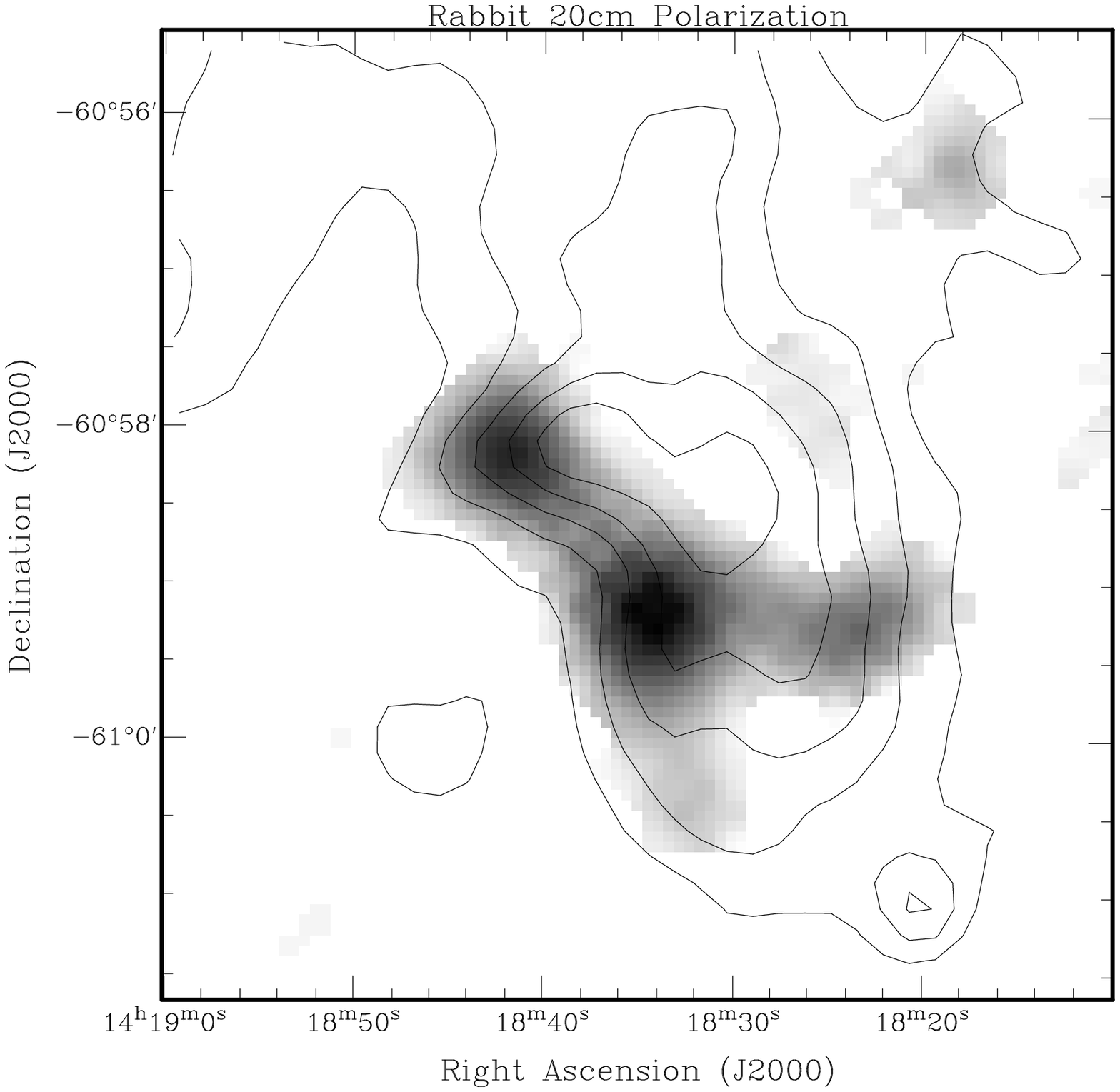, width=0.4\hsize}
            \epsfig{file=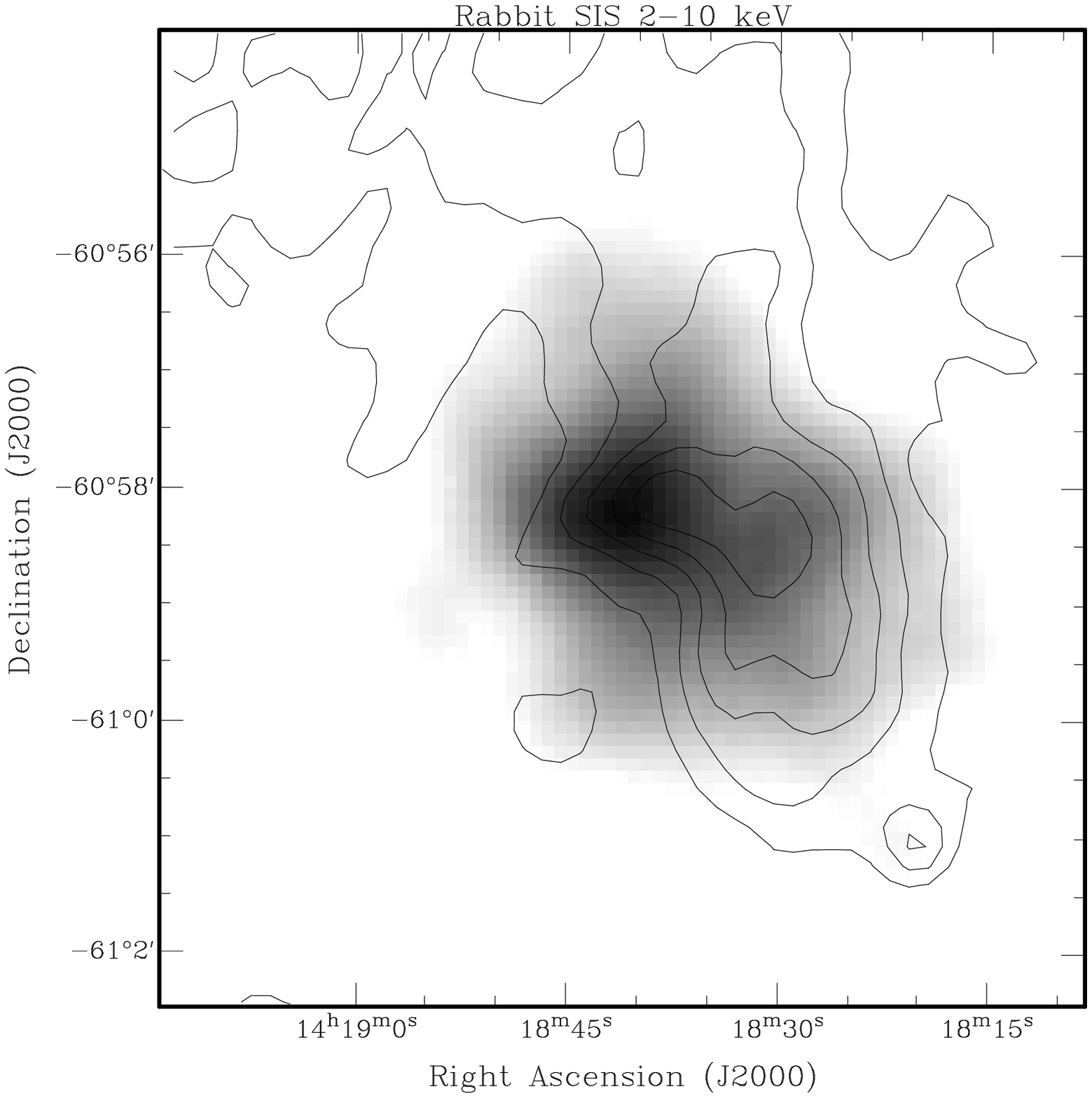,    width=0.4\hsize}}
\caption{{\bf Left:} 20 cm polarization image
of Rabbit nebula, with radio continuum contours. {\bf Right:}
ASCA SIS image of Rabbit. \label{rabpol}}
\end{figure}

The Rabbit is spectrally distinct from the
shell, and a significant source of polarized flux (\cite{r99}). The 
morphology of the polarized flux is doubly peaked (Figure~\ref{rabpol}). 
Spectral tomography of the Rabbit suggests
the polarized region has a spectral index of $\alpha_r \sim -0.3$,
typical of a PWN. The brighter extended X-ray source is coincident
with the Rabbit, and in the SIS image seems to contain a point source
at the location of the upper polarization
peak. The X-ray spectrum and morphology is also suggestive of a PWN, although
detailed spatial structure is impossible to determine with the broad
PSF of {\it ASCA}. 

The Parkes Multibeam Survey recently discovered a very energetic young
($\dot E\sim 10^{37}$ ergs/s) pulsar, PSR J1420$-$6048, 
coincident with the second extended
X-ray source in the upper wing of the Kookaburra (\cite{d01}). 
Folding the X-ray data at the radio period yields a marginal detection
of an X-ray pulse. The detection is somewhat supported by the
pulsed flux being consistent with the {\it ASCA} PSF, while the unpulsed
flux has a significantly broader radial profile (\cite{r01}). While
several of the young objects in the region are at a distance of $\sim 2$kpc, 
the dispersion measure and X-ray absorption  towards the pulsar suggest
a greater distance of $\sim 8$ kpc. Such a distance may be
problematic for an ID as the GeV source, although it should 
be noted that dispersion measure distances can be quite unreliable in 
complex regions such as this one.

The upper wing of the Kookaburra has a strange rectangular shape
reminiscent of SNR 3C 397 (\cite{dr99}). There is no associated infrared flux
seen by IRAS at 60 microns or the MSX mission at 8.3 microns. 
This, along with a suggestion of excess polarization in the ATCA
radio maps, suggests a SNR ID. However, the radio spectral 
index measurements of the entire wing 
are currently too ambiguous to be certain,
although the region immediately around PSR J1420$-$6048 is consistent with
there being a small radio PWN (\cite{r99}).

\subsubsection{GeV J1809$-$2327; A New PWN}

\begin{figure}[h!]
\centerline{\epsfig{file=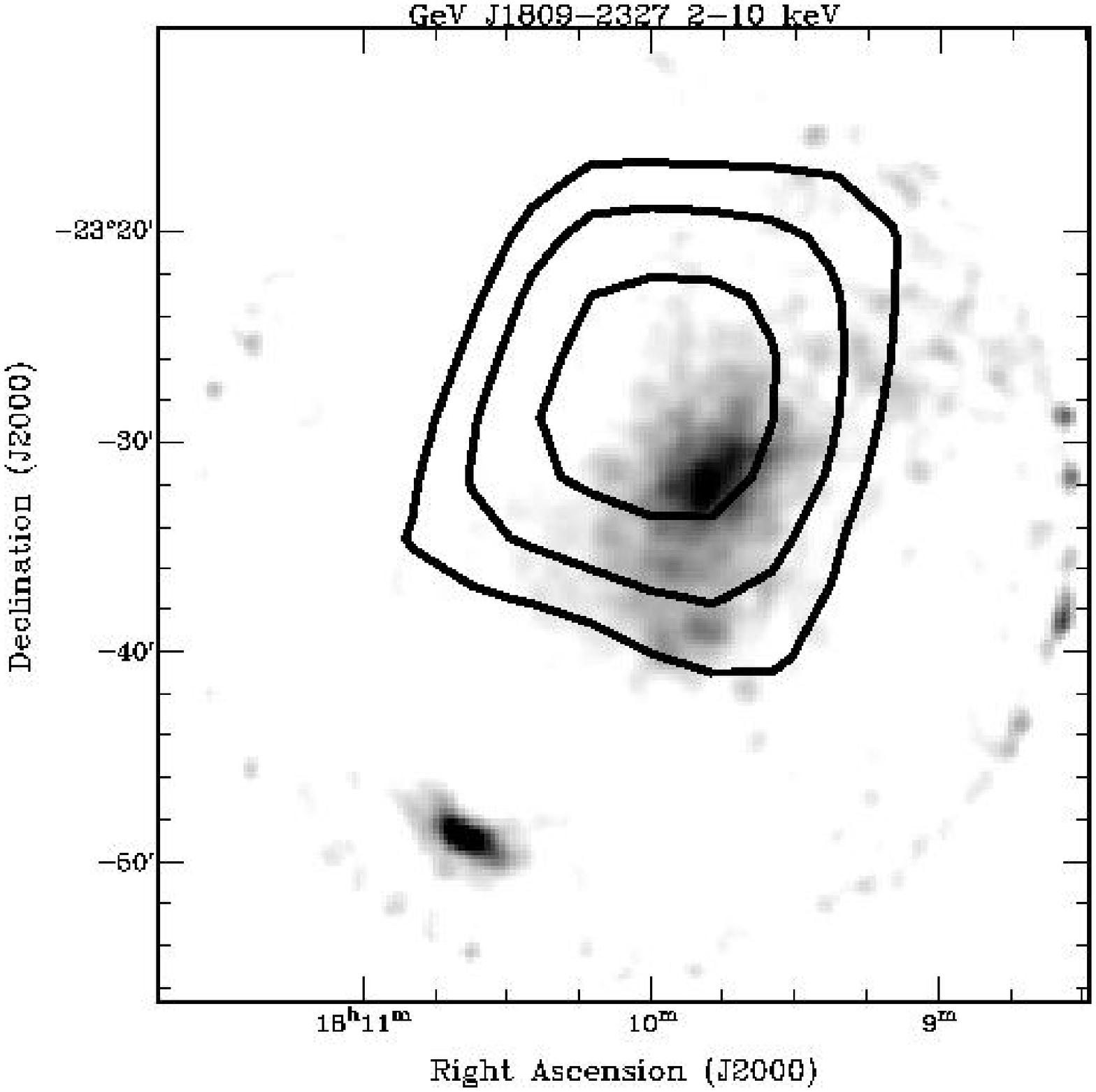, width=0.4\hsize}
            \epsfig{file=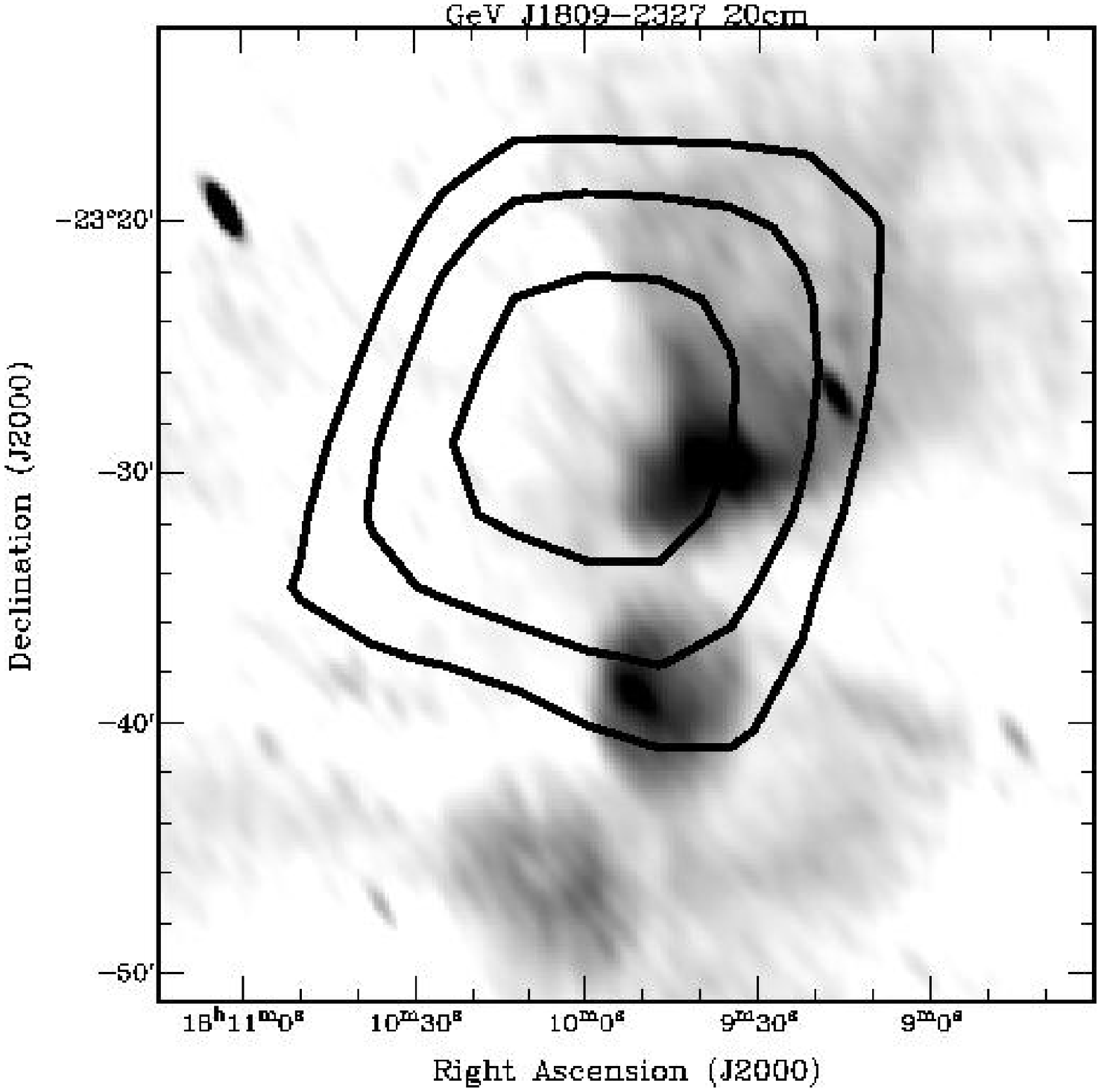,    width=0.4\hsize}}
\caption{{\bf Left:} $ASCA$ 2--10~keV image of GeV J1809$-$2327. 
{\bf Right:}
20 cm VLA image of GeV J1809$-$2327. The funnel
shaped nebula is the likely PWN.  \label{g1809}}
\end{figure}

\begin{figure}[h!]
\centerline{\epsfig{file=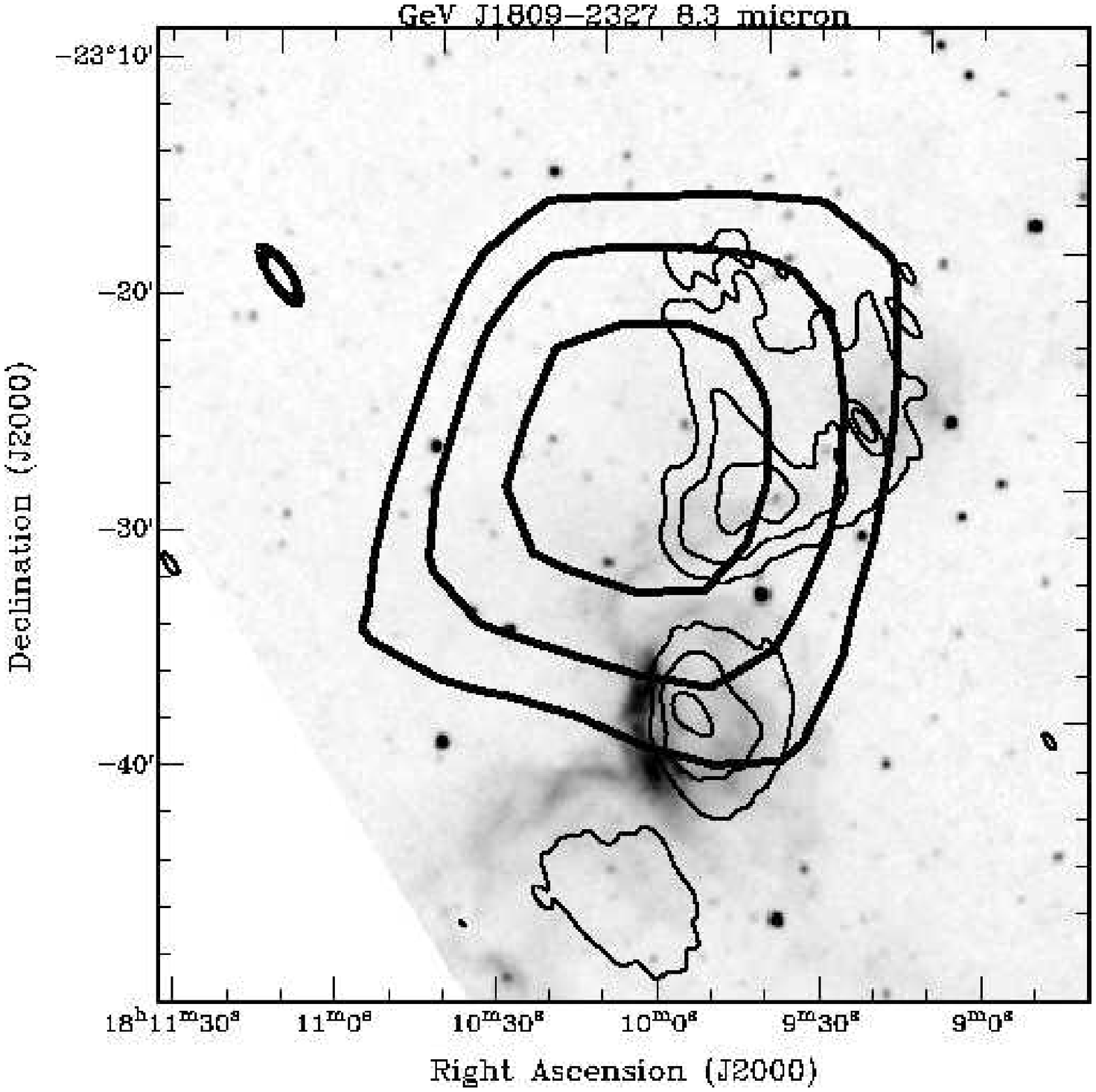, width=0.4\hsize}
            \epsfig{file=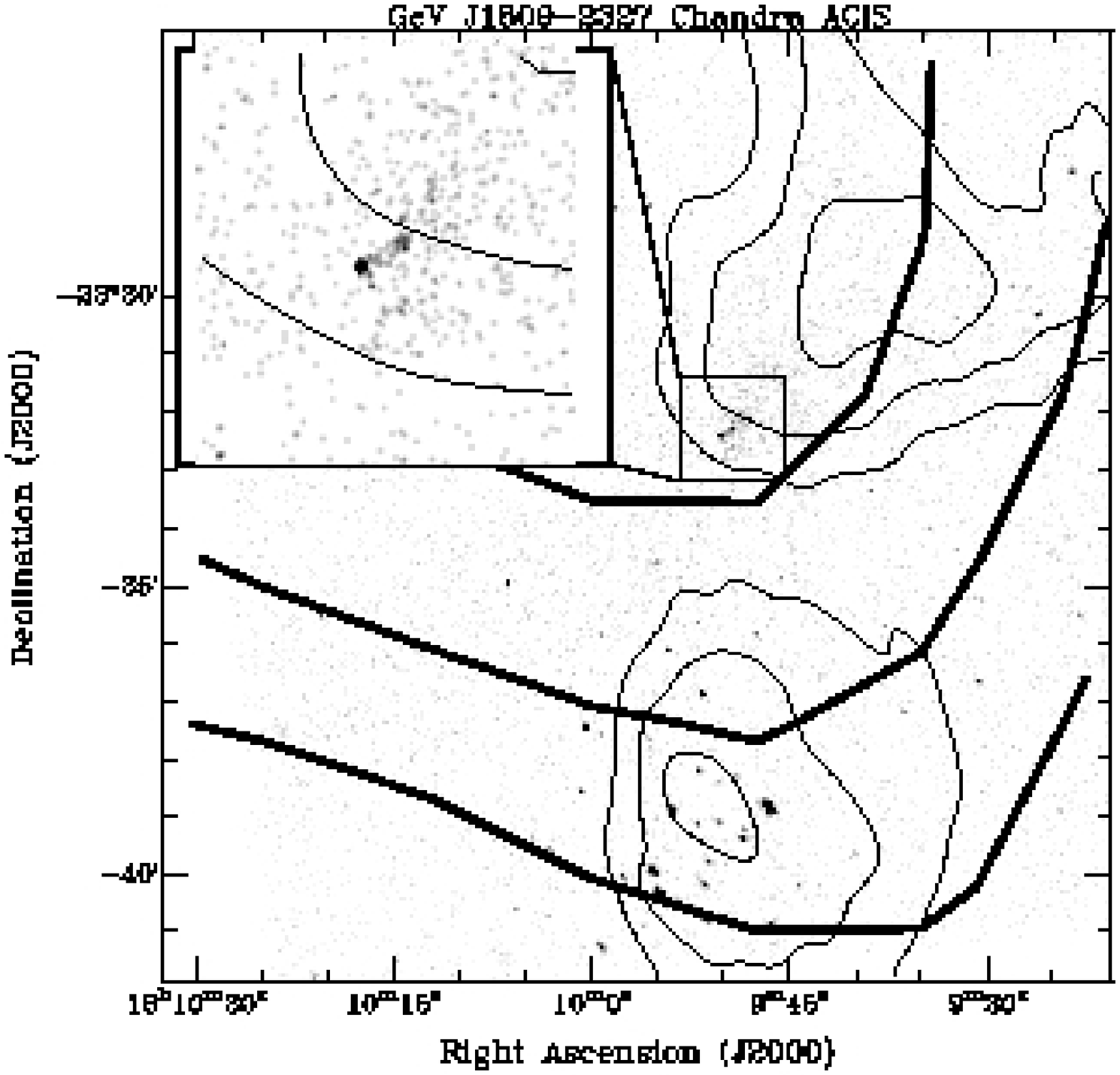,    width=0.4\hsize}}
\caption{{\bf Left:} GeV J1809$-$2327 at 8.3 microns, showing
the thermal nature of the southern radio nebula. The light
contours are of the 20cm radio emission.  {\bf Right:}
Chandra ACIS image of GeV J1809$-$2327. Inset shows pulsar
candidate. \label{acis}}
\end{figure}

The final X-ray moderate, high variability GeV source in the sample
is GeV J1809$-$2327. The {\it ASCA} image (Figure~\ref{g1809}) shows an extended nebula
centrally peaked with several smaller peaks to the south. The
nebula has a power-law spectrum ($\Gamma\sim2.2$), with the
central peak being slightly harder. The smaller peaks are softer and are
coincident with massive young stars. Oka et al. (1999) mapped the
CO emission in the region and noted that the X-ray emission is
surrounded by molecular gas in the Lynds 227 dark nebula. 
20cm and 6cm VLA imaging of the region shows two nebulae
(\cite{rgr01}) consistent with the GeV source. The southern one
seems to be a thermal molecular cloud which is also seen in 
the MSX 8.3 micron image (Figure~\ref{acis}). 
The northern source, coincident with the central
peak of the X-ray nebula, 
has a non-thermal spectrum (energy index $\alpha\sim -0.4$)
and is significantly polarized. These properties, along with 
its shape, suggest a PWN identification.

A short Chandra ACIS image (Figure~\ref{acis}) 
resolves the various point sources (\cite{rr01}), 
and shows that the stellar sources are coincident with the middle,
thermal cloud. The hard central peak of the X-ray nebula is coincident
with a point source at the edge of the PWN candidate. There is
a small trail of emission leading back towards the center of the
radio nebula. This strongly supports the identification of the
X-ray/radio nebula as a PWN. 

%
%
%
%
%

\section{Summary}

X-ray and radio studies of {\it EGRET} error contours have proven 
to be quite fruitful. There are now strong candidate low-energy 
counterparts for the
majority of bright Galactic sources of GeV emission. 
{\it GLAST} should be able
to make positive identifications of most of these sources.
In the particular case of the two
X-ray nebulae in the Kookaburra, {\it GLAST} would be able
to distinguish between them even if both were sources of
GeV emission (\cite{r00}). Although
there are several massive colliding wind binary systems and
SNR coincident with GeV sources, neither source class makes a 
convincing case for copious emission above a GeV. However,
it should be noted that 100 MeV and above error ellipses around
the SNR in this sample are much larger and often shifted from the GeV 
position. There are several other SNR coincident
with sources in the 3rd {\it EGRET} catalog that have significantly steeper 
spectra than the sources in this sample. If SNR are, as a class, 
steep spectrum (photon index $\Gamma > 2$) $\gamma-$ray sources,
then current models of SNR emission and cosmic-ray production
may have to be revised. 

Isolated $\gamma-$ray pulsars (without bright X-ray PWN) seem to be
non-variable, and several new candidates for this older population
of pulsars have been found. Looking at the total numbers of 
isolated pulsars and candidates, roughly half seem to be radio pulsars.
X-ray pulse searches of the radio-quiet candidates with {\it Newton-XMM} 
could prove successful. 

With a strong radio PWN candidate counterpart for GeV J1809$-$2327, 3 of the
4 X-ray moderate, high variability
GeV sources in this flux-limited sample have now been shown to 
contain likely PWN which emit in both 
radio and X-rays
(the other two being the Kookaburra/Rabbit and PSR B1853+01). The fourth,
GeV J1825$-$1310, contains a candidate X-ray PWN which has yet to
be carefully imaged in radio. With the addition of the source coincident
with the likely PWN in CTA 1, also apparently variable, there is
a strong case for synchrotron nebulae being able to produce 
high-energy ($E> 100$ MeV) $\gamma-$rays. 
Noting the apparent soft $\gamma-$ray
variability of the Crab nebula, only the Vela pulsar exists as a
counterexample. 
The variability studies of Tompkins (1999) were sensitive to timescales 
on the order of a few months. This is roughly the synchrotron cooling
timescale of the X-ray emission. The presence of a strongly emitting
$\gamma-$ray nebula would hinder pulse searches of these sources,
since it would significantly reduce the pulse fraction in these sources.
Therefore, even with the greater sensitivity of {\it GLAST}, it may be
necessary to find pulsations at X-ray or radio wavelengths 
before $\gamma-$ray pulsations can be detected in these sources.


\begin{acknowledgments}
We would like to thank V. Kaspi for useful discussions and comments.
MSER acknowledges support from the
Qu\'ebec Merit Fellowship Program. BMG acknowledges the
support of NASA through Hubble Fellowship grant HF-01107.01-98A. 
The National Radio Astronomy Observatory 
Very Large Array is a 
facility of the National Science Foundation operated under
cooperative agreement by Associated Universities, Inc. The Australia
Telescope is funded by the Commonwealth of
Australia for operation as a National Facility managed by CSIRO.
This work made use of several on-line resources, including Skyview and
W3Browse from the High Energy
Astrophysics Science Archive Research Center a service of
the Laboratory for High Energy Astrophysics at NASA/GSFC
and the High Energy Astrophysics Division of the SAO, as well as
the NASA/IPAC Infrared Science Archive.

\end{acknowledgments}


\begin{chapthebibliography}{1}
\bibitem[Baring et al. 1999]{b99}
Baring, M.G., Ellison, D.C., Reynolds, S.P., Grenier, I.A., Goret, P. (1999) {\it ApJ},{\bf 513},311

\bibitem[Berezhko \& V\"olk 2000]{bv00}
Berezhko, E.G., V\"olk, H.J. (2000) {\it ApJ},{\bf 540},923

\bibitem[Case \& Bhattacharya 1999]{cb99} Case, G. Bhattacharya, D. (1999),
{\it ApJ},{\bf 521}, 246

\bibitem[Condon et al. 1991]{c91} Condon, J.J. et al. (1991) {\it AJ},{\bf 102},2041

\bibitem[D'Amico et al. 2001]{d01}
D'Amico, N. et al. (2001) in preparation

\bibitem[de Jager et al. 1996]{dj96}
de Jager, O.C., Harding, A.K., Michelson, P.F., Nel, H.I., Nolan, P.L.,
Sreekumar, P., Thompson, D.J. (1996) {\it ApJ},{\bf 457},253

\bibitem[Dyer \& Reynolds 1999]{dr99}
Dyer, K.K.  Reynolds, S.P. (1999) {\it ApJ},{\bf 526},365

\bibitem[Eichler \& Usov 1993]{eu93}
Eichler, D., Usov, V.(19) {\it ApJ},{\bf 402},271

\bibitem[Frail \& Scharringhausen 1997]{fs97}
Frail, D.A.  Scharringhausen, B.R. (1997) {\it ApJ},{\bf 480},364

\bibitem[Halpern \& Wang 1997]{hw97}
Halpern, J.P.  Wang, F.Y.-H. (1997) {\it ApJ},{\bf 477},905

\bibitem[Harrus, Hughes, \& Helfand 1996]{hhh96}
Harrus, I.M., Hughes, J.P.,  Helfand, D.J. (1996) {\it ApJ},{\bf 464},L161

\bibitem[Harding 2000]{h00}
Harding, A. K. (2000) {\it to be published in High-Energy Gamma-Ray
Astronomy, ed. Aharonian, F.A.,  V\"olk, H.}, astro-ph/0012268

\bibitem[Hartman et al. 1999]{h99}
Hartman, R.C. et al. (1999) {\it ApJS},{\bf 123},79

\bibitem[Hunter et al. 1997]{h97}
Hunter, S.D. (1997) {\it ApJ},{\bf 481},205

\bibitem[Kaspi et al. 1993]{k93}
Kaspi, V.M., Lyne, A.G., Manchester, R.N., Johnston, S., D'Amico, N., 
Shemar, S.L. (1993) {\it ApJL},{\bf 409},L57

\bibitem[Kassim et al. 1989]{k89}
Kassim, N.E., Weiler, K.W., Erickson, W.C., Wilson, T.L. (1989) 
{\it ApJ},{\bf 338},152

\bibitem[Keohane et al. 1997]{k97}
Keohane, J.W., Petre, R., Gotthelf, E.V., 
Ozaki, M., Koyama, K. (1997) {\it ApJ},{\bf 484},350

\bibitem[Keohane et al. 2001]{k01}
Keohane, J.W., Olbert, C.M., Clearfield, C.R., Williams, N.E., 
Frail, D.A. (2001) {\it BAAS}, Late Session 197th AAS Meeting,130.06

\bibitem[Lamb \& Macomb 1997]{lm97}
Lamb, D.Q., Macomb,D.J. (1997) {\it ApJ},{\bf 488},872

\bibitem[McLaughlin et al. 1996]{m96}
McLaughlin, M.A., Mattox, J.R., Cordes,J.M., Thompson, D.J. (1996) 
{\it ApJ},{\bf 473},763

\bibitem[Mirabal et al. 2000]{m00} Mirabal, N., Halpern, J.P., Eracleous, M.,
Becker, R.H. (2000) {\it ApJ},{\bf 541},180

\bibitem[M\"ucke \& Pohl 2001]{mp01}
M\"ucke, A. Pohl, M.  (2001) 
{\it Workshop on Interacting Winds From Massive Stars}, 
ASP Conference Series, Ed. Moffat, A.F.J. \& St-Louis, N. (San
Francisco)

\bibitem[Oka et al. 1999]{o99} Oka, T., Kawai, N., Naito, T., Horiuchi, T.,
Namiki, M., Saito, Y., Romani, R.W., Kifune, T. (1999), {\it ApJ}, 
{\bf 526},764

\bibitem[Punsly 1999]{p99}
Punsly, B. (1999) {\it ApJ},{\bf 516},141

\bibitem[Roberts, Gaensler, \& Romani 2001]{rgr01}
Roberts, M.S.E., Gaensler, B.M., Romani, R.W. (2001) {\it in preparation}

\bibitem[Roberts \& Romani 1998]{rr98}
Roberts, M.S.E., Romani, R.W. (1998) {\it ApJ},{\bf 496},827

\bibitem[Roberts et al. 1999]{r99}
Roberts, M.S.E., Romani, R.W., Johnston, S., Green, A.J. (1999) {\it ApJ},{\bf 515},712

\bibitem[Roberts 2000]{r00}
Roberts, M.S.E. (2000) {\it Ph.D. Thesis, Stanford University}

\bibitem[Roberts, Romani \& Kawai 2001]{rrk01}
Roberts, M.S.E., Romani, R.W., Kawai, N. (2001) {\it ApJS accepted}

\bibitem[Roberts et al. 2001]{r01}
Roberts, M.S.E. et al. (2001) {\it in preparation}

\bibitem[Romani 1996]{r96}
Romani, R.W. (1996) {\it ApJ},{\bf 470},469

\bibitem[Romani et al. 2001]{rr01}
Romani, R.W. et al.  (2001) {\it in preparation}

\bibitem[Sambruna 1997]{sam97}
Sambruna, R. (1997) {\it ApJ},{\bf 487},536

\bibitem[Slane et al. 1997]{s97}
Slane, P., Seward, F.D., Bandiera, R.,
Torii, K., Tsunemi, H. (1997) {\it ApJ},{\bf 485},221

\bibitem[Strickman et al. 1998]{st98}
Strickman, M.S., Tavani, M., Coe, M.J., 
Steele, I.A., Fabregat, J., Marti, J., Paredes, J.M., Ray, P.S. (1998) 
{\it ApJ},{\bf 497},419

\bibitem[Sugizaki et al. 2001]{s01}
Sugizaki, M., Mitsuda, K., Kaneda, H., Matsuzaki, K.,
Yamauchi, S. Koyama, K. (2001) {\it ApJS accepted}, astro-ph/0101093

\bibitem[Tompkins 1999]{t99}
Tompkins, W. (1999) {\it Ph.D. Thesis, Stanford University}

\bibitem[Whiteoak \& Green 1996]{wg96}
Whiteoak, J.B.Z. Green, A.J. (1996) {\it A\&AS},{\bf 118},329
\end{chapthebibliography}
\end{document}